\documentclass[preprint,showpacs,aps,superscriptaddress]{revtex4}
\usepackage{graphicx}
\usepackage{dcolumn}
\usepackage{bm}

\begin{document}
\title{Solvation force for long ranged wall-fluid potentials}
\author{ A. Macio\l ek}  
\affiliation{Institute of Physical Chemistry, 
             Polish Academy of Sciences,\\
        Department III, Kasprzaka 44/52,
            PL-01-224 Warsaw, Poland}

\author{A. Drzewi\'nski}
\affiliation{Institute of  Low Temperature and Structure 
             Research, Polish Academy of Sciences,
      P.O. Box 1410, Wroc\l aw 2, Poland}
\affiliation{Czestochowa University of Technology, Institute of
Mathematics and Computer Science,
ul.Dabrowskiego 73, 42-200 Czestochowa, Poland}
\author{P. Bryk}
\affiliation{Department for the Modeling of Physico-Chemical Processes,
Maria Curie-Sk{\l}odowska University, 20-031 Lublin, Poland\\}
\affiliation{Max-Planck Institut f{\"u}r Metallforschung,
     Heisenbergstrasse 3, D-70569 Stuttgart, Germany\\}
\affiliation{Institut f{\"u}r Theoretische und Angewandte Physik,
       Universit{\"a}t Stuttgart,
     Pfaffenwaldring 57, D-70569 Stuttgart, Germany}
\date{\today}
\begin{abstract}
The solvation force of  a simple fluid confined between identical planar walls 
is studied in two model systems with short ranged fluid-fluid interactions
 and long ranged
wall-fluid potentials decaying as $-Az^{-p}, z\to \infty$, for various values of $p$. Results for the Ising spins system are  obtained
in two dimensions at vanishing
bulk magnetic field $h=0$ by means of the density-matrix renormalization-group method;
results for the truncated Lennard-Jones (LJ) fluid  are obtained within the
nonlocal density functional theory.   At low  temperatures
the solvation force $f_{solv}$ for  the Ising film
is repulsive  and decays for large wall separations
$L$ in the same fashion as the boundary field $f_{solv}\sim L^{-p}$,
whereas for temperatures larger than the bulk critical temperature $f_{solv}$ is  attractive
and the asymptotic decay is $f_{solv}\sim L^{-(p+1)}$.
For  the LJ fluid system  $f_{solv}$
is always repulsive away from the critical region
and decays for large $L$ with the the same power law 
as the wall-fluid potential. We discuss the influence of the critical Casimir effect and of capillary condensation on the behaviour of the solvation force.

\end{abstract}
\baselineskip 9mm
\maketitle
\section{Introduction}
\label{sec:intro}
Fluid  mediated interactions  between two surfaces
or large  particles,  usually referred  to as
 solvation forces $f_{solv}$~\cite{evans:90:0}, may lead to subtle effects
 ~\cite{israelachvili:91:0} that can be relevant  in colloidal systems and
for  many applications and modern technologies such as lubrication,  adhesion or friction.
There is a rapidly growing literature on experimental investigations
and other phenomena associated with  solvation
 forces. Direct measurements with the  Surface Force Apparatus (SFA)~\cite{israelachvili:78:0}
 have revealed the  richness of behaviour  of these forces, for example,
their sensitivity to specific properties  of the intervening fluid.
Dependence of the solvation  force on the
chemical and physical properties of
confining surfaces is also of much interest.
Different surfaces have been developed  and used in SFA by
adsorbing  or depositing
a thin film of some other material on a mica surface, for example, lipid monolayers
 or bilayers, metal films,
polymer films or  other macromolecules such as proteins~\cite{israelachvili:91:0,smith:88:0,parker:88:0,lee:89:0}.

On the other hand, the theory of the solvation forces which would
help to  interpret  measurements performed for different liquids and different  surfaces
 is far from being complete even for simple model systems.
One of the issues which has not been systematically investigated
is  how the properties of the solvation
force $f_{solv}(L)$, such as its sign, magnitude and  asymptotic  dependence
 on the distance between surfaces,
vary with the range and the strength of the  substrate-fluid potential and with
 the thermodynamic state.
In the present paper we address these issues
for a simple  fluid confined between two identical
 parallel solid substrates separated by a distance $L$.

The majority  of results  available  for  the solvation force
in this simple system
comes from  a  model fluid  in which both the fluid-fluid
 interatomic potential
and the wall-fluid potential  $V_s({\bf r})$ are of short range.
Here, by $f_{solv}(L)$ we are referring only to the fluid contribution
to the force, the direct wall-wall contribution is not included.
Let us summarize briefly results that are the most relevant
 for the present paper.
The precise form of the solvation force depends
on  the details of $V_s({\bf r})$, the bulk state point, as well as on $L$.
Theory and computer simulations  show  that at small wall separations
the solvation force exhibits oscillatory behaviour
 with $L$.
These  oscillations, observed  in  direct measurements of the solvation force in real fluids,  reflect  packing effects
 that produce oscillations in the density profile  for liquids  near walls
 ~\cite{evans:90:1}.
The behaviour  of  $f_{solv}(L)$ reflects the behaviour of the density
profile  also in  the limit of large separation, $L\to \infty$.
General statistical mechanical arguments predict
that far from  the bulk critical temperature $T_c$ and from any phase transitions
(condensation phenomena)   $f_{solv}(L)$ for $L\to \infty$  must decay  to zero
 in the same fashion as the profile at a single wall decays  to its
bulk value, i.e.,  as the radial
distribution  function $g(r)$
 of the bulk fluid~\cite{evans:92:0,henderson:92:0,evans:93:0,evans:94:1}.
Thus, for  $V_s({\bf r})$ of  finite range and $L\to \infty$,
$f_{solv}(L) \sim e^{-\alpha_0L}$
or,  for sufficiently high densities and/or temperatures,
(states on the oscillatory side of the  so-called Fisher-Widom line)
$f_{solv}(L)\sim exp(-\alpha_0L)\cos (\alpha_1L)$.
The  quantities $\alpha_0$ and $\alpha_1$ depend on only the bulk pair correlation
function.
The sign of the solvation force and its temperature dependence at fixed $L$
are available  from  mean-field
 (lattice, Landau, density functional)
analyses~\cite{evans:86:0,marconi:88:0,parry:92:0,krech:97:0,dietrich:98:0,upton:99:0}
and from exact transfer matrix methods for two-dimensional strips~\cite{evans:94:0}.
For simple fluids confined between identical walls $f_{solv}(L)<0$ for large $L$, i.e.,
the net force between the substrates is attractive for large separations.
Away from the bulk critical temperature $T_c$,  $f_{solv}(T)$ is small in magnitude (weakly attractive).
On approaching $T_c$ at vanishing  ordering field, $h=0$, where
 $h\sim \mu -\mu _{sat}$,
 where $\mu _{sat}$ is the chemical potential at coexistence,
the force increases in magnitude and exhibits a minimum. For strongly adsorbing walls 
this minimum is  located above $T_c$ 
and  occurs for $L$ of the order of the bulk correlation length $\xi _b$.
 Critical scaling arguments predict~\cite{fdg:78:0,fss} that at bulk criticality $f_{solv}(L)$ decays algebraically for large separations, i.e.,
\begin{equation}
\label{eq:Casf}
f_{solv}(L)\sim k_BT_cA_{12}(d-1)L^{-d}~~~~~~~~~~~~~L\to \infty.
\end{equation}
$d$ is the  spatial dimension and $A_{12}$ a universal number (Casimir amplitude)
for $2\le d\le d_>$, where the upper critical dimension $d_>=4$ for the Ising
universality class. $A_{12}$ is negative for identical walls, i.e. the Casimir force is attractive.

In the present paper we study two  model systems  in which 
fluid-fluid interactions are short ranged   but
the substrate-fluid
potential $V_s({\bf r})$  is  {\it  long ranged},
 and enquire   how the features of the solvation force described
above change with the range  and the strength of $V_s({\bf r})$.
As a first system  we consider 
a  (Ising) lattice gas in a slit geometry subject to identical boundary fields
which depend on the distance $l$ of a lattice site from the boundary 
\begin{equation}
\label{eq:bf}
V_s({\bf r})\equiv H^s_l=\frac{h_1}{l^p}
\end{equation}
with $h_1, p\ge 0$.
The Ising model is particularly useful in fundamental studies of  finite-size and
surface effects in confined fluids  since exact calculations are available,
at least in two dimensions and for short
 ranged boundary
 fields~\cite{evans:94:0,au-yang:75:0,abraham:71:0,abraham:73:0,maciolek:96:0,maciolek:99:0}. 
Another advantage  is that 
 in $d=2$ this model is  amenable to the
systematic  investigation for {\it  arbitrary boundary and bulk fields}
 by means of   density-matrix renormalization-group
 (DMRG) method~\cite{clsys}. 
 The DMRG method,  based on the transfer matrix approach,
provides a very efficient algorithm for constructing 
 the effective transfer matrices for large systems. Comparisons with exact results for the case of vanishing bulk magnetic field and 
boundary fields acting only on spins in the surface layers  show that
this technique gives   very accurate results  in a wide range of temperatures,
including  near the bulk critical temperature,
 and for large widths of the  strip~\cite{maciolek:99:0,carlon:98:0,drzew:00:1}.
Here,  using the  DMRG method 
we calculate $f_{solv}$  as a function of temperature along the bulk two-phase
coexistence line, i.e., for   vanishing bulk magnetic field,
for different  values of $p$, $h_1$ and $L$. We also study the asymptotic decay,
with $L$, of the solvation force for strongly adsorbing walls (large $h_1$) and
temperatures away from $T_c$.

Lattice models of a confined fluid cannot describe
 packing effects that 
are reflected in the oscillations of the solvation force at small wall separations. The other failing of the (Ising) lattice gas model
  is  that it has an exact
 particle-hole symmetry. For real fluids  such symmetry is only approximate
which can  be of relevance  for the properties of $f_{solv}$
 away from criticality. 
Therefore, to make our analysis  more complete we also study a Lennard-Jones (LJ) fluid in  a slit geometry.
Specifically, we consider  a truncated  LJ 12-6 fluid-fluid 
intermolecular pair potential and a wall-fluid potential of the  form:
\begin{equation}
\label{eq:wp}
V_s({\bf r})\equiv V_s(z)=4\varepsilon_{fw}\left[ \frac{2}{15}\left(\frac{\sigma}{z}\right)^{9}
-\left(\frac{\sigma}{z}\right)^p\right]\;,
\end{equation}
where $p=2,3$; $\sigma$ is the diameter of the fluid species, while $\varepsilon_{fw}$
describes the strength of the wall--fluid interactions.
We note that for $p=3$,  $~V_s(z)$ models a wall  which is assumed
to comprise a half space of LJ particles~\cite{israelachvili:91:0}.
We calculate $f_{solv}$
using a   nonlocal density functional theory (DFT)
along a  similar thermodynamic path as in the Ising system, i.e.,
  as a function of the temperature along the  bulk two phase-coexistence line
and at the critical density for $T>T_c$.
Our results refer to the  densities  on the liquid side of this line. 
We also investigate  the asymptotic 
dependence of the solvation force on the distance
 between the walls at fixed temperature away from
$T_c$.

The asymptotic behaviour of the solvation force
 for the Lennard-Jones fluid  follows
 from the analysis 
by Attard {\it et al.}~\cite{attard:91:0}
based  on  the wall-particle Ornstein-Zernike (OZ) equations.
 Using the
hypernetted-chain closure  these authors  derived the interaction
 free energy per unit area between 
the planar walls $F_{00}(L)$
as a convolution of wall-solvent pair-correlation functions.
For a power-law fluid-fluid  interaction potential
$-Ar^{-n}, n>3,  r\to \infty$  and a  wall-fluid
 potential decaying as $-Bz^{-p}$ for $z\to \infty$
a  formula  for the  asymptotic behaviour of  $F_{00}(L)$  (Eq.(6.11)
 of Ref.\cite{attard:91:0}) is
\begin{equation}
\label{eq:attard}
F_{00}(L)\sim  \beta u_{00}(L)+\frac{2\rho B}{p-1}L^{1-p}
-\frac{2\pi\rho^2A}{(n-2)(n-3)(n-4)}L^{4-n},~~~~~L\to\infty
\end{equation}
where $u_{00}(L)$ is the  direct wall-wall interaction potential  per unit area
and $\rho $ is the density of the  fluid.
From Eq.~(\ref{eq:attard})  it follows that
in the case of truncated LJ fluid, when the power-law tails are omitted, the behaviour of
 $f_{solv}(L)=-\left(\partial F_{00}(L)/\partial L\right)_{T,\mu}$ for large $L$
is
\begin{equation}
\label{eq:LJasym}
f_{solv}(L)\sim \frac{2\rho  B}{L^p},~~~~~~~~~~~~L\to \infty,
\end{equation}
where  $B=4\varepsilon_{fw}$ and we have neglected the contribution
due to the direct wall-wall interaction potential.
Thus the solvation force is repulsive and decays with the same power law
 as the wall-fluid potential.
The above prediction is treated  as a reference point for the
 analysis of the asymptotic behaviour of our results.
We find  that, as one  expects,  Eq.~(\ref{eq:LJasym})
is indeed valid {\it provided} one is away from the critical temperature
 and from any phase transitions.
The influence of the critical Casimir effect and of  capillary condensation
on the behaviour of the solvation force is also discussed in the present paper. 

It is reasonable to expect that away from $T_c$ and from
 any phase transitions
the 'magnetic' analog of the solvation force 
in  system of Ising spins subject to identical boundary fields
 should have the same asymptotic decay power law  as
 the boundary field $H^s_l$.
Somewhat surprisingly, this is not the case  for temperatures far above $T_c$ where
we find  that  $f_{solv}(L)$ decays as $L^{-(p+1)}$ for $L\to \infty$.
For low temperatures the decay agrees with the prediction (\ref{eq:LJasym}).
In order to understand the nature of this particular   behaviour we analyse the
appropriate Landau theory.

The  paper is organized as follows. 
Section~\ref{sec:isres} is devoted to the Ising model studies.
We  define the model and  briefly review the known results pertinent 
 to the present studies. We then proceed to describe  the results of the 
DMRG studies.
In Sec.~\ref{sec:landau}  a continuum Landau theory is
 investigated for  long ranged  surface fields.
In Sec.~\ref{sec:dft}  DFT calculations are presented and discussed. 
Section ~\ref{sec:sum} summarizes our results and makes some  conclusions.

\section{Ising model results}
\label{sec:isres}
\subsection{The model}
\label{subsec:model}
We consider an Ising ferromagnet in a slit  geometry subject to identical 
boundary fields.
Our DMRG results refer to the $d=2$ strip defined on the square lattice of the size $L\times M, M\to\infty$.
The lattice consists of $L$ parallel rows at spacing $a=1$, so that the width of the strip is $La=L$. We label successive rows by the index $l$.
At each site, labeled $(l,k)$, there is an Ising spin variable taking the value
$\sigma _{lk}=\pm 1$.
We assume nearest-neighbour  interactions of strength $J$
 and Hamiltonian of the form
\begin{equation}
\label{eq:Ham}
{\cal H}=-J\{\sum_{<lk,l'k'>}\sigma_{lk}\sigma_{l'k'}+\sum_{l=1}^LH_l\sum_k\sigma_{lk}\}.
\end{equation}
The first term in (\ref{eq:Ham}) is a sum over all nearest-neighbor pairs of
sites while in the second term $H_l=H^s_l+H^s_{L+1-l}$ is the total boundary  magnetic field experienced by a spin in row $l$.
The single-boundary field  $H^s_l$ is assumed to have a form:
\begin{equation}
\label{eq:bf1}
H^s_l=\frac{h_1}{l^p}
\end{equation}
with $p >0$, and the reduced amplitude of the boundary field  $h_1\ge 0$.

\subsection{Solvation force for short ranged boundary fields}
\label{subsec:surfield}
All previous work on the behaviour  of the solvation force 
in  Ising films  used localized boundary fields acting only on spins in surface layers $l=1$ and $l=L$: 
\begin{equation}
\label{eq:srf}
H_l=h_1\delta _{1,l}+h_2\delta _{L,l}.
\end{equation}
In the limit $h_1, h_2\to \infty$ Eq.~(\ref{eq:srf}) corresponds  to the $(++)$
fixed spin boundary conditions, widely studied in the literature. 

The total excess  free energy per unit area for the case of
  identical  surface 
fields $h_1=h_2$ and vanishing bulk  magnetic field $h=0$ 
  can be written  as
\begin{equation}
\label{eq:fren}
f_{ex}(L)\equiv L(f(L,T,h_1)-f_b(T))+2f_w(T,h_1)+f^*(L,T,h_1)
\end{equation}
where $f$ is the total free energy per site, $f_b$ is the bulk free energy,
 $f_w$ is the $L$-independent
surface excess free energy contribution from each surface,
 and $f^*$ is the finite-size contribution to the free energy.
 All energies are measured in units of $J$ and the
temperature in units of $J/k_B$.
$f^*$, which vanishes for $L\to \infty$,  gives rise to the 
generalized force,  which is analogous  to the solvation force 
between the walls in the case of confined fluids~\cite{evans:90:0},
\begin{equation}
\label{eq:solfordef}
f_{solv}=-(\partial f_{ex}(L)/\partial L)_{T,h_1}.
\end{equation} 
For  identical surface
 fields the solvation force is attractive for
all thermodynamic states, i.e., $f_{solv}<0$.
The asymptotic decay of the solvation force  depends on
 the temperature range.
For temperatures sufficiently far away from the bulk critical temperature $T_c$,
  $f_{solv}$  decays as $\exp(-L/\xi _b)$,
where $\xi _b$ is the bulk correlation length~\cite{evans:94:0}. 
Near $T_c$,  $f_{solv}$ becomes long ranged   as a result
 of critical fluctuations ~\cite{fdg:78:0},
 a phenomenon which is known as the  critical Casimir effect~\cite{krech:99:0}.
At $T=T_c$ and $h=0$ the asymptotic decay is given by Eq.~(\ref{eq:Casf}).
For the case of $(++)$ boundary conditions the temperature dependence of the solvation force at fixed $L$ and $h=0$, or equivalently the scaling 
 function $W_{++}({\tilde y})$ is  
defined by 
\begin{equation}
\label{eq:fssCas}
f_{solv}/k_BT_c\equiv (d-1) L^{-d}W_{++}({\tilde y}),
\end{equation}
  where the scaling variable  ${\tilde y}\equiv \tau(L/\xi_0)^{1/\nu}$.
 In $d=2$ the scaling function 
was determined by exact transfer matrix methods~\cite{evans:94:0}.
Here $\tau\equiv (T-T_c)/T_c$ and $\xi=\xi_0\tau^{-\nu}$ (with $\nu=1$ in 
$d=2$) is the  bulk correlation length.
It was found that  $ f_{solv}$ plotted   as a 
function of temperature
attains a pronounced minimum 
{\it  above} the bulk critical temperature $T_c$   
 when ${\tilde y}\equiv \tau(L/\xi^+_0)^{1/\nu}=2.23$, or $L\sim 2.23 \xi$.
The amplitude of the correlation length above $T_c$ is 
 $\xi^+_0\approx 0.5673$.

It was also shown~\cite{evans:94:0} that the scaling function has the property, for $d=2$,
\begin{equation}
\label{eq:scpr}
W_{++}({\tilde y})=W_{00}(-{\tilde y}),
\end{equation}
where subscript $00$ refers to $h_1=h_2=0$.
This implies that at $h=0$ and fixed $L$ the function
$f_{solv}(T)$ evaluated  for free boundaries  has  its minimum  {\it below} $T_c$.
The behaviour  of the function $f_{solv}(T)$ in the crossover between  
$h_1=\infty$ and $h_1= 0$  has not been studied. In the next section we
perform such  an investigation  for both  short ranged  and 
 long ranged boundary fields.

Finally, we note that for free boundaries $h_1=h_2=0$  the location
 of the minimum 
of the solvation force, at $h=0$,  is  associated with the 
critical temperature, $T_{c,L}$, which for $d\ge 3$ and large 
 but finite $L$  denotes the  end of  the
two-phase coexistence ~\cite{evans:94:0,maciolek:01:0}. 
$T_{c,L}$ lies on the  $h=0$ axis
 and  is shifted {\it below}  the temperature of the
 bulk critical point $T_c$ by an amount given by
 the expression~\cite{fisher:81:0},  $\tau\sim -L^{-1/\nu}$.
For nonvanishing surface fields, $h_1=h_2>0$, the situation is different.
In this  case the preferential adsorption of $(+)$ spins at each wall
leads to a shift  of the bulk phase boundary  in the $(h,T)$ plane to $h<0$.
 This phenomenon of capillary condensation strongly influences  properties
 of the Ising films -  also  above the (capillary)
critical point $(h_{c,L}(h_1),T_{c,L}(h_1))<T_c$
~\cite{maciolek:01:0,frank:03:1}.
The  minimum  of the scaling function $W_{++}({\tilde y})$  lies
{\it above} $T_c$ which can be accounted for by  the fact that
the most  pronounced features  in the solvation force
  occur along the continuation to higher $T$ of
the capillary condensation  line  $h_{co}(T)$~\cite{maciolek:01:0}.
The same is true  for
  other thermodynamic  and structural quantities that
arise in the film  geometry, such as the  specific heat, adsorption, or
 longitudinal correlation length $\xi _{\parallel}$.
Specifically,
$f_{solv}$ at fixed $T>T_{c,L}$ has a deep minimum at some $h<0$, which
corresponds  roughly to the continuation of the line
 $h_{co}(T)$. As the temperature increases the minimum
approaches $h=0$ and
 decreases in depth.
The stronger  $h_1$, the bigger is the shift of  the capillary critical point
from the bulk coexistence line and the  further
 above $T_c$ reaches the line along which  the most pronounced minima
of the solvation  force occur.
The shape of the scaling function $W_{++}({\tilde y})$ reflects this  behaviour.
 $W_{++}({\tilde y})$ is very weak for $T<T_c$  when  the condensation
line is far away from  $h=0$,  and develops
the minimum above $T_c$ when the minima of the solvation force
that occurs along the
 continuation of $h_{co}(T)$ approach the line  $h=0$.

\subsection{DMRG results for long ranged boundary fields}
\label{subsec:DMRG}
The density-matrix renormalization group (DMRG) method  was introduced in 1992
by White as a numerical  algorithm to study ground-state properties of
quantum-spin chains \cite{white}. In spite of the name, the method has
only some
analogies with the traditional renormalization group being essentially
the numerical, iterative basis, truncation method. Later, the DMRG was
adapted by Nishino for two-dimensional classical systems at non-zero
temperatures~\cite{clsys}. It is particularly well suited to study systems in
 confined geometry since it  deals naturally  with lattices of a size
$L \times \infty$. Generally, the DMRG method works best with open
boundary conditions, which makes the technique appropriate to take into
consideration the effects of surfaces.
We have implemented  a finite-size version of the DMRG algorithm designed
for accurate studies of finite-size systems \cite{white}.
For a comprehensive review of a background, achievements, and limitations
of the method, see Ref.\cite{dmrg}.

In a transfer matrix approach a leading
eigenvalue $\lambda_L$ of a transfer matrix $T_L$
\begin{equation}
\label{eq:dmrg1}
T_L |v_L\rangle = \lambda_L |v_L\rangle,
\end{equation}
gives a free energy per spin of an Ising strip as
\begin{equation}
\label{eq:dmrg2}
\beta f(L)=-\frac{1}{L} \ln \lambda_L .
\end{equation}
The components  of the eigenvector $|v_L\rangle$  related to the leading eigenvalue give
probabilities of various configurations.
For classical spin systems the DMRG method is based on the transfer-matrix
approach, where the leading eigenvalue and its eigenvector of
the effective transfer matrix are calculated 
numerically. This method can be  employed 
for a number of problems for which no exact solutions are available
(e.g. Ising systems in a presence of  bulk magnetic  field, or, as in the 
present case, subject to  long ranged surface fields).

The main idea of the DMRG technique is to avoid the proliferation of
states when the size of the system grows. Generally, a  number of 
configurations in a
Hilbert space of an  Ising strip grows very fast with its width $L$
(as $2^L$). Therefore, it is  practicaly impossible  to solve exactly 
systems with $L>25$. In the DMRG approach one eliminates  the least
probable (in the density matrix sense) states and keep only the most
important ones.  From this
stage   calculations are  not exact anymore, but we can obtain a very 
efficient approach if the weights of the discarded states are very small.
Starting with a small system (e.g. $L=4$  in our case), 
for which $T_L$ can be diagonalized exactly, one adds
iteratively two spin rows  in the middle of a strip until the allowed 
(in the computational sense)
size of   effective matrices is reached. Then further addition of new 
spins forces one to discard simultaneously
the least important states to keep the size of  effective matrices
fixed. This truncation is done through the construction of the
reduced density matrix whose eigenstates provide an  optimal basis set.
The size of the effective matrix is then substantially smaller than 
the original dimensionality of the configurational space \cite{bulk2}
$(2m)^2 \ll 2^L$.
 Generally, the larger is $m$,  the better is the  accuracy.
 In the present case we keep this parameter up to $m=50$. 
It is worth mentioning that at low temperatures in order to renormalize
a  transfer matrix we have to use  its
two eigenvectors corresponding to phases with the  opposite magnetization.
To construct the reduced density matrix from the lowest eigenstates
one has to diagonalize an  effective transfer matrix $T_L$ at each DMRG step.
Therefore, we used the so-called Arnoldi method \cite{arnoldi}. 

To calculate the solvation force we proceed in the same way as 
in the case of the short ranged  surface fields. We calculate
 the excess free energy per unit area $ f_{ex}(L)\equiv 
\left(f-f_b\right)L$ at $L_0+2$ and
 $ L_0$. For vanishing bulk magnetic field $f_b$ 
is known exactly~\cite{onsager:45:0}.
Having values 
$f^{ex}(L_0+2)$ and 
$f^{ex}(L_0)$ we approximate the derivative in Eq.~(\ref{eq:solfordef})
 by a finite difference
$f_{solv}=- (1/2)(f^{ex}(L_0+2)- f^{ex}(L_0))$.

\begin{figure}
\includegraphics[width=16cm,clip]{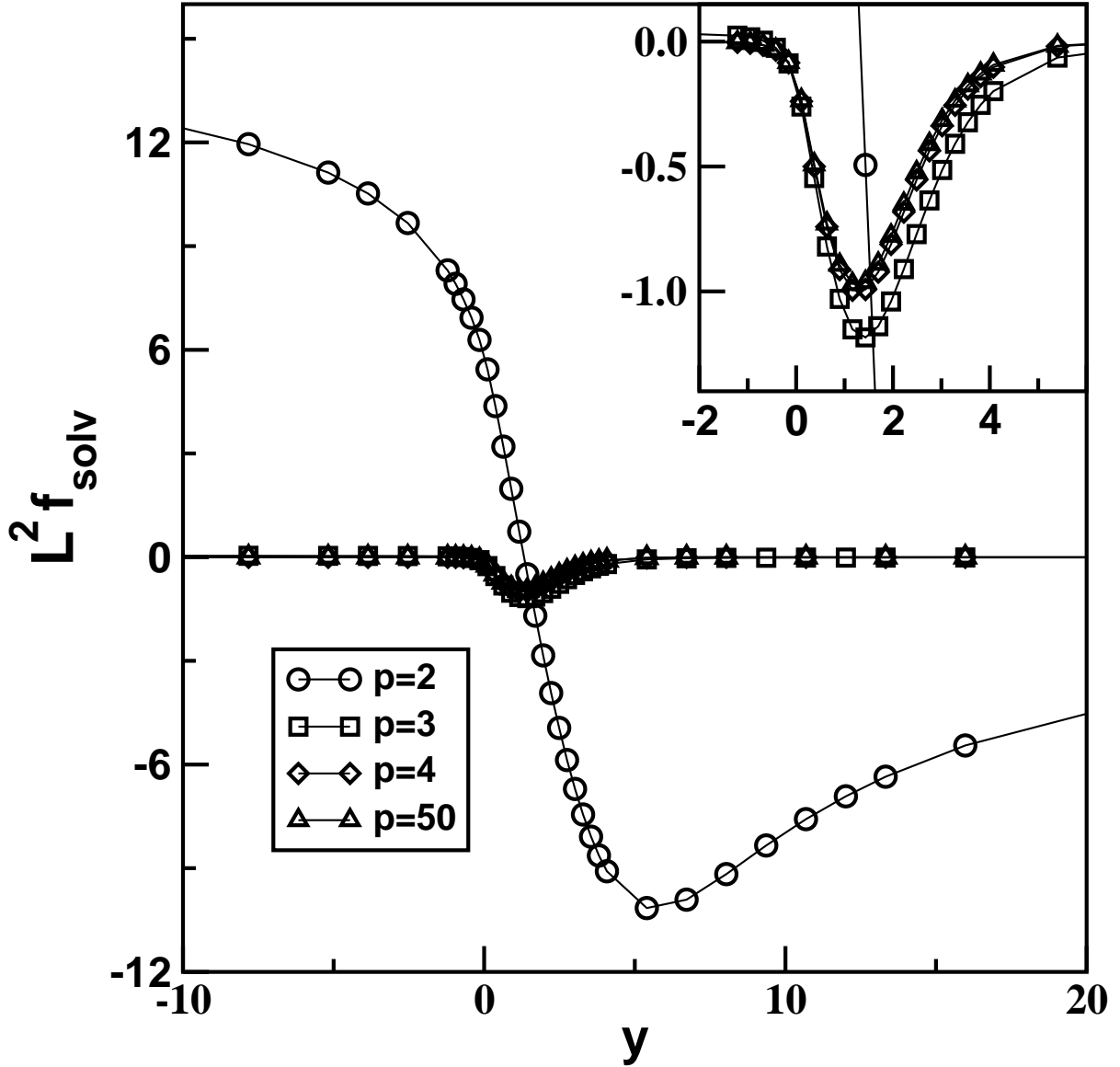}
\caption{\label{fig:fsT}
Solvation force (in units of $J$) for
 an
 Ising strip of the  width $L=300$ at vanishing bulk magnetic field  subject to
 long ranged  boundary  fields $H_l=H^s_l+H^s_{L+1-l}$, $H^s_l=h_1/l^p$
 as a function of the scaling variable $y\equiv \tau (L)^{1/\nu}=\tau L$
 for various values of  $p$
  and the amplitude $h_1\approx 8.16$ calculated using DMRG method.
  The inset shows, on an expanded scale, that results for $p=50$ and $p=4$
  differ only very slightly in the near-critical region.
   For all values of $p$ $~f_{solv} $ is attractive above $T_c$ and, except
   for $p=50$,  repulsive
   below $T_c$.
}
\end{figure}

We have chosen  four different power-laws 
 describing the decay of the surface 
fields, i.e., we consider  Eq.~(\ref{eq:bf}) with   $p=50, 4, 3, 2$.
The behaviour of the system with  $p=50$  should be close to the one with the
 short ranged  surface fields.
 $p=4$ corresponds to  dispersion forces in $d=2$.
For each  type of  boundary  field  we calculate  the solvation force as a 
function of the temperature $f_{solv}(T)$ along the
bulk coexistence line $h=0$ for a range of amplitudes  $h_1$.
Results  for a fixed  strip  width  $L=300$, and fixed field  
strength parameter   $h_1\approx 8.16$
 are plotted as a function of  the scaling
 variable $y\equiv  \tau L^{1/\nu}$ (see Eq.~(\ref{eq:fssCas})) in Fig.~\ref{fig:fsT}.
For this value of $h_1$ the system with short ranged surface fields is almost
identical  to the infinite surface field scaling limit, i.e., to  $(++)$
boundary conditions.
The  inset shows a magnified plot of the region around the
bulk  critical temperature $T_c\approx 2.269185$. 
We have found that, as expected, for $p=50$ 
the solvation force behaves as for the case of  short ranged
surface fields, i.e., it is always negative, exhibits a minimum 
located  above $T_c$ for $y\approx 2.23$,
 and away from the critical region   approaches zero from below.
For  all other values of $p$  we observe that  $f_{solv}$ 
 is {\it repulsive} below $T_c$.   As $T$ approaches $T_c$
the critical fluctuations give  rise to the  Casimir effect.
The fluctuation-induced Casimir force is always  attractive and we can see that
 $f_{solv}$  changes its  sign to become  negative as the temperature
increases towards $T_c$. This effect becomes  stronger
on increasing the range of the surface fields (see Fig.~\ref{fig:fsT}). 
Strikingly, for high temperatures $f_{solv}$ remains {\it attractive}
and  approaches zero from below
{\it independently} of the value of $p$.
Notice, that for $p=2$   the solvation force 
is distinctly stronger than for the higher values of $p$.  

\begin{figure}
\includegraphics[width=16cm,clip]{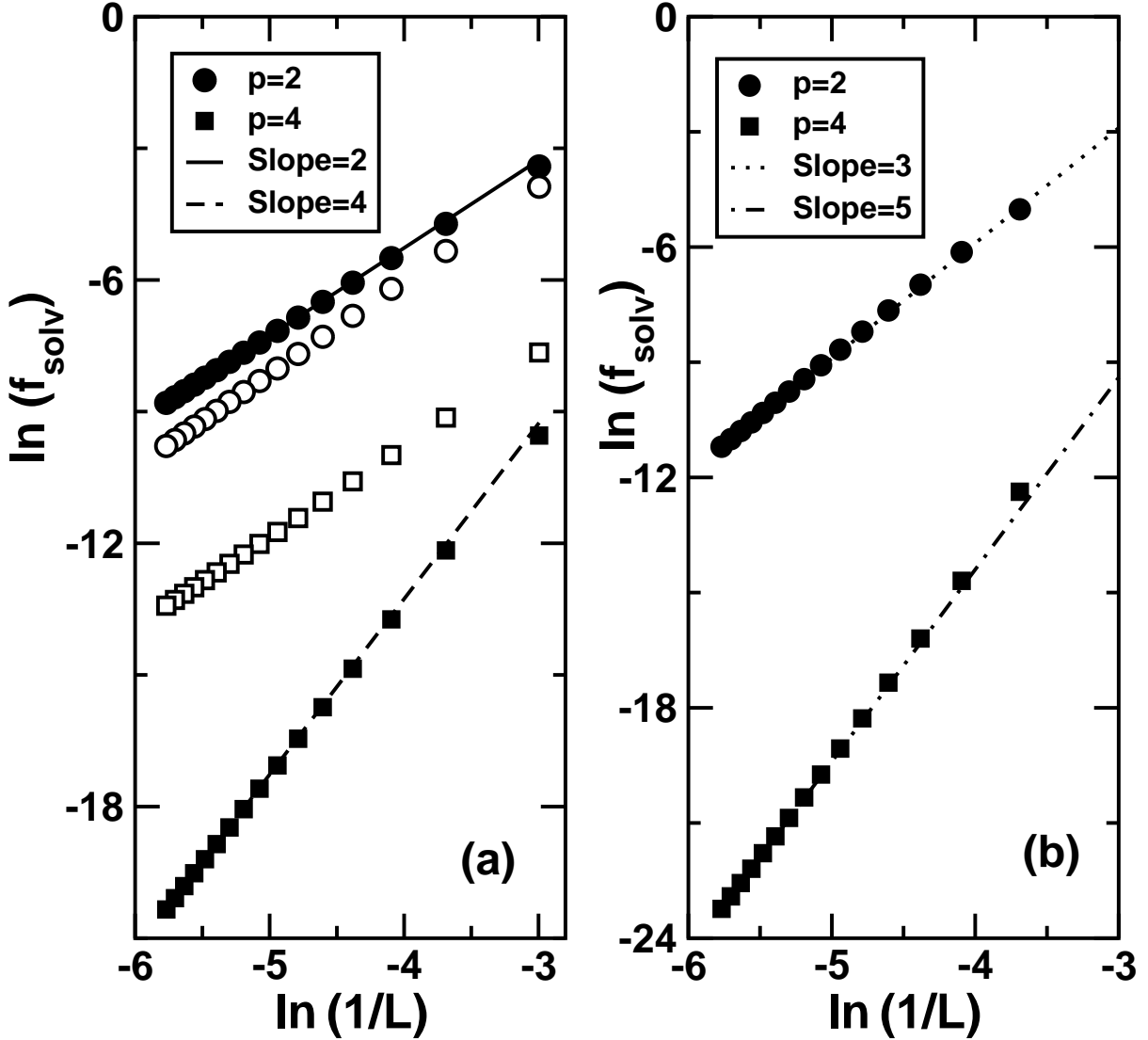}
\caption{\label{fig:r3}
Log-log plot of the solvation force (in units of $J$)
versus $1/L$  for two values of the
power $p$ describing the  decay  of the boundary field $H^s_l$, $p=2$ and 4.
({\bf a}) displays DMRG results for $T=0.79 T_c$ (filled symbols)  and $T=T_c$
(unfilled  symbols).  Straight lines correspond to $f_{solv}(L)\sim  L^{-p}$.
Data at $T_c$ align parallel to the line of a slope 2 showing the result
 $f_{solv}(L)\sim  L^{-d}$ (see Eq.~(1)) with $d=2$.
({\bf b}) displays DMRG results for $T=1.23T_c$. Straight lines correspond to $f_{solv}(L)\sim  L^{-(p+1)}$.
 }
\end{figure}

Turning now to the asymptotic behaviour of the solvation force
 we   plot in Fig.~\ref{fig:r3}  $\ln f_{solv}$ versus $\ln (1/L)$
for various   $L$ between 20 and 320. The force has been calculated
for  fixed   $h_1\approx 8.16$ at 
three different temperatures, corresponding to  $T\ll T_c$,  $T=T_c$, and $T\gg T_c$. 
For $p=$50 we
find asymptotic  decay typical of that for  short ranged surface potentials, i.e., 
away from $T_c$ the solvation force decays exponentially with $L$.
In Fig.~\ref{fig:r3} we  display only results for the parameter $p=$ 4 and 2
but for $p=$3  we observe the similar behaviour.
Filled symbols in Fig~\ref{fig:r3}a  represent results calculated at $T/T_c=0.79$,
circles for $p=2$ and squares for $p=4$.
The straight lines  in this figure   have slopes equal to  $p$
(the solid line has a slope 2 and the  dashed line has a slope 4) and fit  the
data at this  temperature,  far below the critical temperature.
Symbols  in   Fig~\ref{fig:r3}b represent results calculated at $T/T_c=1.23$.
The straight lines in  this figure  have slopes equal to  $p+1$ 
(the dotted line has a slope 3 and the dot-dashed line has a slope 5).
 Symbols not connected by lines in
Fig~\ref{fig:r3}a  are   results obtained
 at the bulk critical
temperature.
The critical scaling analysis predicts  that if the  long ranged
boundary field decays sufficiently rapidly, i.e.,
 when $(1/2)(d+2-\eta)-p$ is negative, the boundary field 
is irrelevant in the RG sense~\cite{diehl}.  $\eta $ is a critical exponent
of a spin-spin correlation function. 
The correction to the leading short-ranged behaviour 
given by (\ref{eq:fssCas}) should be of the order of $L^{-(d-2+\eta)/2-p}$
and may become dominant for $L\gg \xi_b\log \xi_b$~\cite{dantchev}.
 In the present case of the $d=2$ Ising model
$\eta $ is  equal to 0.25 so that for $p\ge 2$ we expect to see 
the power law $L^{-2}$
for the asymptotic decay of the solvation (Casimir) force at 
$T=T_c$  (see Eq.~(\ref{eq:Casf})). 
 We have  found  very good agreement with this prediction  
 for  all values of $p$. One can see in Fig.~\ref{fig:r3}a that  
both circles and squares  corresponding to $p=$ 2 and 4, respectively,
  align almost  parallel to a straight  line of  slope 2. 
We have checked that for $p=4$ the data fit a
  line with a slope $\simeq 2.02$, 
whereas for $p=2$ the fitted line has  a slope $\simeq 2.14$.
Notice that for $p=2$ the  correction to the leading
 decay of the solvation force  is 
of the order of $L^{-2.125}$.
In order to observe a better limiting  behaviour for $p=2$ 
it would be necessary to go to much  larger values of $ L$.
Another useful check of  the irrelevance of the long ranged boundary fields
is the value of the Casimir amplitude $A_{++}$, a universal quantity defined as
$A_{11}(d-1)=W_{11}(0)$ (see Eq.~(\ref{eq:fssCas}). Its value is equal to 
$-\pi/48\approx 0.065$ in $d=2$ Ising model with  $(++)$ boundary conditions ~\cite{blote:86:0} and should be the same for all $p\ge 2$. For each value of $p$,  $h_1\approx 8.16$ and $L=300$
we have calculated  the Casimir amplitude and found a very good agreement
with the prediction. For example, 
 $A_{++}\approx 0.065$ for $p=50$,   $\approx 0.067$ for $p=4$, and $\approx 0.072$ for $p=3$.
 Again for $p=2$ corrections to the finite-size scaling  become important and 
we observe the significant  deviation from the
universal value, i.e. $A_{++}\approx 2.5$.

The above results show that the  asymptotic
 behaviour of the solvation force for the Ising
system  below $T_c$ agrees with the formula   
(\ref{eq:LJasym}) obtained by Attard et al~\cite{attard:91:0} for fluids, 
i.e., for large $L$,  $~f_{solv}(L)$ decays
with the same power-law as the boundary field.
In order to enquire  about the amplitude,
 in  Fig.~\ref{fig:r4} we  plot  $f_{solv}$  calculated for $L=100$ and $h_1\approx 8.16$,
scaled with its value at $T/T_c=0.75$ as a function of 
 the reduced temperature $T/T_c$,
 i.e. $f^{\ast}_{solv}\equiv f_{solv}(T/T_c;h_1,L,p)/f_{solv}(0.75;h_1,L,p)$. 
It  follows from this figure, that  for $T<T_c$ the amplitude 
of the power-law is the same for all $p$
and  depends weakly on the temperature. We have checked 
that its value  is  equal to $2m^*(T)h_1$, where $m^*(T)$ is the  bulk
 spontaneous magnetization. Thus, for large $L$,  $f_{solv}\sim 2m^*(T)h_1/L^p$ and hence $f^{\ast}_{solv}(T/T_c)\sim m^*(T/T_c)/m^*(0.75)$.
 Note however, 
$m^*=0$ above $T_c$, and in this temperature range we observed a different power-law decay.
Moreover, the force becomes  attractive
which is in disagreement with (\ref{eq:LJasym}).
In the next Section we investigate the origin of this  behaviour
within a continuum  Landau theory.

It is also instructive to observe the effect of changing the amplitude $h_1$
on the temperature dependence of the solvation force for various range
of the surface potential.
Figs.~\ref{fig:fsTh_1} a,b display  $f_{solv}$ as a function
of the variable $y$ at fixed $L=100$ along
 the bulk coexistence line $h=0$ for a  selection of the strength
parameter $h_1$  and  for $p=50$ and 2, respectively.  
As the amplitude $h_1$ varies between zero and 8.16
 we observe a nontrivial 
crossover behaviour of the solvation force in the critical regime,
associated with the change of
 the  position  of the  minimum   from the temperature
below $T_c$ to the temperature above $T_c$.
Note that for short ranged boundary fields $h_1=0$ corresponds to
 the ordinary transition universality class,
whereas $h_1\approx 8.16$ corresponds to the normal
 transition universality class.
In the crossover between these two universality classes  a  scaling variable
$L/l_1$ where $l_1\sim  h_1^{-\nu/\Delta_1}$ becomes relevant. $L/l_1\ll 1$ 
corresponds to  the ordinary
transition and $L/l_1\gg 1$ corresponds to  the normal transition. For $L/l_1=O(1)$
a strong deviations from the universal behaviour are expected.
 In $d=2$ Ising model $l_1\sim h_1^{-2}$. For  a long ranged boundary
 fields there appear an additional  scaling variable $h_1L^{(d+2-\eta)/2-p}$ which, as already mentioned above, is irrelevant in the RG sense  but gives corrections to the
finite size scaling which may be important for small 
wall separations $L$~\cite{dantchev}.
In Fig.~\ref{fig:fsTh_1} we can see that as $h_1$ increases from zero the 
minimum located below $T_c$ reduces and shifts towards $T_c$.
 At a certain
small  value  of 
$h_1$ a second shallow  minimum  appears on the other side of  $T_c$. 
 At some small  value of $h_1$, which depends on $p$,
the  minima become symmetric. For $p=50$, which is almost a short ranged
boundary field,  it takes place
at $h_1\approx 0.0605$, i.e. when $L/l_1=O(1)$. As  $h_1$ increases  further the minimum below $T_c$
diminishes  and a  minimum above $T_c$  becomes deeper;
finally a single minimum above $T_c$ remains.
The formation of a two maxima structure in the function $f_{solv}(T)$
is also observed  by varying 
the distance between the walls  at fixed small value of $h_1$.
Examination of the shapes  of the  functions  of Fig.~\ref{fig:fsTh_1} reveals
that the symmetry  between $f_{solv}(T)$  for the  free and $(++)$
boundary conditions   given by Eq.~(\ref{eq:scpr}) 
 is broken for the long ranged surface fields.
This may be connected with the presence of a new
 scaling variable $h_1L^{(d+2-\eta)/2-p}$.
Clearly the solvation force  for $h_1=0$ does not change with
the parameter  $p$, whereas
for large values  of $h_1$  the depth of the 
 minimum of $f_{solv}(T)$ increases 
 and its position 
moves monotonically towards  higher
values of $T$ as the range of the surface potential increases.

The  change of the locus  of the minimum  of the solvation force 
as the parameter $p$ is varied
is consistent with the behaviour of 
 the specific heat $C_H(L,T;p,h_1)=-T(\partial^2f/\partial T^2)_L$
 and of  longitudinal spin-spin  correlation length
 $\xi _{\parallel}$. 
These quantities  are 
 readily calculable from the total free
energy obtained in DMRG.
$\xi _{\parallel}$  
 may be expressed in terms of the ratio of the largest $\Lambda _0$
and second largest $\Lambda _1$ eigenvalues of the transfer matrix
$\xi^{-1}_{\parallel} (L,T;p,h_1)=-\ln[\lambda _1/\lambda _0]$.
We have  calculated  $C_H$ and  $\xi^{-1}_{\parallel}$
  as a function
of  temperature for $L=100$,  $h_1\approx 8.16$ and various $p$.
Both quantities exhibit extrema  above $T_c$ which decrease  and shift
 towards  higher
values of $T$ as the range of the surface potential
 increases (see Fig.~\ref{fig:rysCvXi} a, b).
If the position of the extrema of $f_{solv}(T)$, 
$C_H(T)$, and  $\xi^{-1}_{\parallel}(T)$ at $h=0$ 
is governed by 
 capillary condensation~\cite{maciolek:01:0}
 (see Sec.~\ref{sec:isres}), then their
shift  towards higher values of $T$ indicates that
the (capillary) critical point  moves further away from the bulk
 coexistence line $h=0$ as the range of the surface potential increases.

\section{Analysis of Landau functional}
\label{sec:landau}
In this section we analyse the Landau theory  corresponding to the Ising model 
considered  in Sec.~\ref{sec:isres} in order to
understand why at  high  temperatures the solvation force is negative and, for large
wall separation $L$,
decays as $L^{-(p+1)}$,  whereas 
at low temperatures $f_{solv}(L)$  is positive and the form of decay agrees with
 the prediction (\ref{eq:LJasym}), i.e.  $f_{solv}(L)\sim L^{-p},~~~L\to\infty$.

In the Landau theory
the magnetization profile $m(z)$
in  a slit of width $L$ subject to the boundary field  $H(z)$ 
is obtained by minimizing  the free-energy functional
 (per unit area and per $k_BT$)~\cite{fisher:81:0}:
\begin{equation}
\label{eq:lan1}
F[m]=\int_0^Ldz\left[\frac{1}{2}b\left(\frac{dm}{dz}\right)^2+f_b(m(z))-H(z)m(z)\right]
\end{equation}
where $b$ is a positive constant.
$f_b(m)$ is the  bulk free-energy density 
\begin{equation}
\label{eq:lan2}
f_b(m)=\frac{1}{2}a\tau m^2+\frac{1}{4}am^4-hm,
\end{equation}
where $\tau$ is the reduced deviation from the bulk critical temperature,
 $a$ is constant
 and $h$  is the bulk magnetic field.
We consider a long ranged  boundary field $H(z)$ 
of the  form
\begin{equation}
\label{eq:lan3}
H(z)=h_1\left[ \frac{1}{(z+\lambda)^p}+\frac{1}{(L+\lambda-z)^p}\right],
\end{equation}
where we have introduced the parameter $\lambda$ satisfying
 $L\gg \lambda\gtrsim 1$ in order 
to assure  that  $H(z)$ is well behaved  at the boundaries. 
In the final analysis  we take the limit $\lambda/L\to 0$.

In order to obtain  the solvation force one  has to find
 the equilibrium magnetization profile $m(z)_{eq}$ and then calculate 
the total excess free energy per unit area 
\begin{equation}
\label{eq:excess}
f_{ex}(L)\equiv F(L)-Lf_b(m_b),
\end{equation}
where $F(L)\equiv F[m_{eq}]$ and $m_b$ is the bulk magnetization
at given $T,h$.
The solvation force is defined  by  Eq.~(\ref{eq:solfordef}).

Henceforward, since we are interested in the asymptotic decay of $f_{solv}(L)$
for temperatures above $T_c$ and at vanishing bulk magnetic field, for our analysis it is sufficient 
to  keep only the term quadratic in $m$ in the expression
for the bulk free energy density.
In this temperature range the bulk correlation length
 $\xi _b=(b/a\tau)^{1/2}$ and  $m_b=m^*=0$.

Minimization of Eq.~(\ref{eq:lan1})
yields an Euler-Lagrange
equation
\begin{equation}
\label{eq:lan4}
\frac{d^2m(z)}{dz^2}=\xi_b^{-2}m(z)-(1/b)H(z).
\end{equation}
We assume the following  boundary conditions
\begin{equation}
\label{eq:lan5}
m(0)=m(L)=1,
\end{equation}
valid for the considered  case,
i.e. large $h_1$,  strongly adsorbing walls.
Imposing the condition that at equilibrium the profiles are symmetric
around $z=L/2$ yields further
\begin{equation}
\label{eq:lan5a}
\frac{dm}{dz}_{z=L/2}=0.
\end{equation}
The solution to the Euler-Lagrange equation (\ref{eq:lan4})
is equal to the sum of the general solution, $m_{g}(z)$, 
of the corresponding homogeneous differential equation (with $H(z)=0$), 
and any particular solution, $m_{p}(z)$, of the inhomogeneous equation (with $H(z)\ne 0$):
\begin{equation}
\label{eq:lan6}
m(z)=m_{g}(z)+m_{p}(z).
\end{equation}
Above $T_c$ the general solution  of  equation (\ref{eq:lan4}) 
with $H(z)=0$
satisfying  the symmetry condition (\ref{eq:lan5a}) is
\begin{equation}
\label{eq:lan7}
m_{g}(z)=A\left[ e^{-z/\xi _b}+ e^{-(L-z)/\xi_b}\right].
\end{equation}
The required  solution  of the inhomogeneous  equation can be expressed
in the form of a  power series in $H(z)$ 
\begin{equation}
\label{eq:lan8}
m_p(z)= B_1\left[ \frac{1}{(z+\lambda)^p}+\frac{1}{(L+\lambda-z)^p}\right]
+B_2\left[ \frac{1}{(z+\lambda)^{p+2}}+\frac{1}{(L+\lambda-z)^{p+2}}\right]+\ldots
\end{equation}
where the constants $B_1, B_2,\cdots $ have  to be determined
 by equating coefficients. 
We  substitute  (\ref{eq:lan8}) for $m(z)$ and its second derivative
in the differential equation (\ref{eq:lan4}) and  notice that 
for  large separation $L\to \infty$
the lhs of this equation is subdominant to the rhs. 
Therefore, in the limit $L\to \infty$, we can approximate $m_p(z)$ 
by the solution of  the equation
\begin{equation}
\label{eq:lan9}
0=\xi_b^{-2}m(z)-(1/b)H(z),
\end{equation}
i.e., $m_p(z)\approx (\xi_b^2/b)H(z)$.
Thus the general solution of  (\ref{eq:lan4}) is
\begin{equation}
\label{eq:lan10}
m(z)\approx A\left[ e^{-z/\xi_b}+ e^{-(L-z)/\xi_b}\right]+\frac{\xi_b^2}{b}H(z),
\end{equation}
where the constant  $A$ can be determined from the boundary conditions
(\ref{eq:lan5}).
Substituting the above solution into the functional (\ref{eq:lan1}), 
performing the integral over $z$ and  then the derivative with 
 respect to $L$, one can
 find the asymptotic behaviour of $f_{solv}(L)$ when  $L\to \infty$.
Notice, that for  $T>T_c$  the bulk spontaneous magnetization $m^*=0$, 
so that  $f_b(m_b)=0$ in the Eq.~(\ref{eq:excess})
 for the excess free energy per unit area.

Substitution of the equilibrium magnetization profile (\ref{eq:lan10}) 
 into  the integrand in
 (\ref{eq:lan1}) yields  terms purely
 exponentially  and purely algebraically decaying with  $z$,   as well as 
terms in which an  exponential  and an  algebraic  decay mixes together. There  are no terms
decaying in the same fashion as the boundary field. Such terms, if they were to exist,
   would yield the power-law decay of $f_{solv}$ that is consisitent with the result by Attard {\it et al.} (Eq.~(\ref{eq:LJasym})).
Careful analysis reveals that the  leading asymptotic  behavior  of
the solvation force arises from a purely algebraically decaying
term in  $(b/2\xi_b^{2})m^2(z)-H(z)m(z)$, namely
\begin{equation}
\label{eq:lan103}
-\frac{\xi_b^2h_1^2}{b}\frac{1}{(\lambda+z)^{p}(L+\lambda-z)^p}.
\end{equation}
The contribution to the $f_{solv}$ from the above term is (see Appendix)
\begin{equation}
\label{eq:lan14}
\frac{\xi_b^2h_1^2}{b}\frac{\partial}{\partial L}\int_0^Ldz\frac{1}{(\lambda+z)^{p}(L+\lambda-z)^p}=-\frac{2\xi_b^2h_1^2}{b\lambda^{p-1}}\frac{p}{p-1}\frac{1}{L^{p+1}}+O(L^{-(p+2)}),~~~L\to\infty.
\end{equation}
Other purely algebraically decaying terms in 
 $(b/2\xi_b^{2})m^2(z)-H(z)m(z)$ give contributions to
 the solvation force that are  $O(L^{-(2p+1)})$ and the 
 mixed terms cancel out.

Contributions to the solvation force arising from 
  $(1/2)b\left(dm/dz\right)^2$ decay faster than $L^{-(p+1)}$
 for  $L\to \infty$.
In this expression there are four mixed terms:
\begin{equation}
\label{eq:lan11}
\frac{ e^{-z/\xi_b}}{(z+\lambda)^{p+1}};~~~~~~~~~~~~~~~~\frac{e^{-(L-z)/\xi_b}}
{(L+\lambda-z)^{p+1}}.
\end{equation}
\begin{equation}
\label{eq:lan111}
\frac{ e^{-(L-z)/\xi_b}}{(z+\lambda)^{p+1}};~~~~~~~~~~~~~~~~\frac{ e^{-z/\xi_b}}
{(L+\lambda-z)^{p+1}}.
\end{equation}
The asymptotic behaviour 
of contributions to the solvation force arising from the  above terms is
(see Appendix)
\begin{equation}
\label{eq:lanl12}
-\frac{\partial}{\partial L}\int_0^Ldz\frac{e^{-z/\xi_b}}{(L+\lambda-z)^{p+1}}=
-\frac{\partial}{\partial L}\int_0^Ldz\frac{e^{-(L-z)/\xi_b}}{(\lambda+z)^{p+1}}=
\frac{(p+1)\xi_b}{(L+\lambda)^{p+2}}+O(L^{-(p+3)}),~~L\to \infty
\end{equation}
and 
\begin{equation}
\label{eq:lanl12a}
-\frac{\partial}{\partial L}\int_0^Ldz\frac{e^{-z/\xi_b}}{(\lambda+z)^{p+1}}=
-\frac{\partial}{\partial L}\int_0^Ldz\frac{e^{-(L-z)/\xi_b}}{(L+\lambda-z)^{p+1}}=-\frac{e^{-L/\xi_b}}{(\lambda+L)^{p+1}}+\frac{1}{\lambda ^{(p+1)}},~~L\to \infty.
\end{equation}
 Purely  algebraically decaying terms in the  expression $-(b/2)m(z)\left(d^2m/dz^2\right)$
give contributions $O(L^{-(p+2)})$. 

In conclusion, we have found that for $T>T_c$
\begin{equation}
\label{eq:asfs}
f_{solv}(L)=-\frac{2\xi_b^2h_1^2}{b\lambda^{p-1}}\frac{p}{p-1}\frac{1}{L^{p+1}}+O(L^{-(p+2)}),~~~L\to\infty.
\end{equation}
Note that since $b>0$ the solvation force is negative.
This asymptotic behaviour does not change if we take into account higher order
terms in the solution for $m_p(z)$ (see Eq.~(\ref{eq:lan8})).

The fact that in the limit $L\to\infty$  the solvation force decays 
 faster than 
the boundary field is  due to the absence in the expression for the
 free energy density of terms proportional to $H(z)$.
Below $T_c$ the solution of the Euler-Lagrange equation (Eq. (\ref{eq:lan4})
with an additional term $\tau^{-1}\xi_b^{-2}m^3(z)$ in the rhs )
for large $L$ 
has the  form $m(z)=m^{\ast}+{\tilde m}(z)$, where $m^{\ast}\ne 0$
is the  bulk spontaneous magnetization. Similarly to the case of $T>T_c$, the  function
${\tilde m}(z)$ can be expressed  as a sum of the  solution
of the corresponding homogeneous differential equation (with $H(z)=0$), 
$m_{h}(z)$, 
and a solution of the inhomogeneous equation (with $H(z)\ne 0$),
$m_{inh}(z)$, i.e. ${\tilde m}(z)=m_{h}(z)+m_{inh}(z)$ .
For large $L$ the approximation  (\ref{eq:lan10})
 for $m_{inh}(z)$ is still valid,
therefore in the expression for the free energy we can expect terms
$\sim m^{\ast}H(z)$ which then yield an  asymptotic decay of the solvation force of the same
 form as that of the boundary field.

\section{Density functional theory results}
\label{sec:dft}
The model considered in this Section is a van der Waals fluid of the bulk density
$\rho_{b}$ confined in a slit of width $L$. Each of the walls interacts with the fluid via
the potential described by Eq.~(\ref{eq:wp}) with $p=2,3$.
The total external potential of the system $V_{ext}(z)$ is a sum of the contributions
from both walls, $V_{ext}(z)=V_{s}(z)+V_{s}(L-z)$.
The fluid particles interact via the standard Lennard--Jones potential
\begin{equation}
u(r)=\left\{\begin{array}{ll}
4\varepsilon_{ff}\left[ \left(\frac{\sigma}{r}\right)^{12}
-\left(\frac{\sigma}{r}\right)^{6}\right]\;, & r<r_{cut}\\
0\;\;, &  r>r_{cut}
\end{array}
\right.
\end{equation}
where $\varepsilon_{ff}$ describes the strength of the fluid--fluid interactions and
$r_{cut}$ is the cut-off distance. We set  $r_{cut}=2.5\sigma$.
The system is studied by means of a density functional theory (DFT) \cite{Evans92}.
Within this approach the grand potential $\Omega$ of the system is a functional of
the local density $\rho({\bf r})$          
\begin{equation}
\Omega[\rho]=F[\rho]+\int d^3r\, \rho({\bf r})(V_{ext}({\bf r})-\mu)   
\end{equation}
where $\mu$ is the  chemical potential of the fluid.                              
The free energy functional $F$ is a sum of two parts, $F=F_{id}+F_{ex}$.
The ideal gas contribution is known exactly  
\begin{equation}
\beta F_{id}=\int d^3r\,[\ln(\Lambda^3\rho({\bf r})-1]\rho({\bf r})\;,          
\end{equation}
where $\beta=(k_BT)^{-1}$ and $\Lambda$ is the de Broglie wavelength.
The excess (over ideal) free energy is a sum of  reference hard--sphere $F_{ex}^{(hs)}$
and attractive $F_{ex}^{(att)}$ contributions. The latter is evaluated in a mean--field fashion
\begin{equation}
F_{ex}^{(att)}=\frac{1}{2}\int d^3r\int d^3r'\,\rho({\bf r})\rho({\bf r}')
u_{WCA}(|{\bf r}-{\bf r}'|)\;
\end{equation}
where $u_{WCA}(r)$ corresponds to  the Weeks-Chandler-Andersen \cite{Weeks79}
division of the interparticle potential
\begin{equation}
u_{WCA}(r)=\left\{
\begin{array}{ll}
-\varepsilon_{ff}\;, & r < 2^{\frac{1}{6}}\sigma\\
u(r)\;, & r>2^{\frac{1}{6}}\sigma
\end{array}
\right.
\end{equation}
The reference hard--sphere part of the excess free energy is evaluated
within the framework
of the Fundamental Measure theory (FMT)
of Rosenfeld \cite{Rosenfeld89,Rosenfeld93}
\begin{equation}
\beta F_{ex}^{(hs)}=\int d^{3}r \;\Phi(\{n_{\alpha}\}) \;,
\end{equation}
where $n_{\alpha}$ denote weighted densities
\begin{equation}
n_{\alpha}({\bf r})=\int d^{3}r' \rho({\bf r}')~
w_{\alpha}({\bf r}-{\bf r}') \;,
\end{equation}
with six different geometrical weight functions $w^{(\alpha)}$  (four scalar and two vector--like
\cite{Rosenfeld93}).
There are several expressions for the excess free energy density $\Phi$.
We have chosen for the present problem the original Rosenfeld functional, where
the excess free energy density is given by
\begin{equation}
\Phi(\{n_{\alpha}\})=-n_{0} \ln (1-n_{3})+\frac{n_{1}n_{2}-{\bf n}_{1}\cdot
{\bf n}_{2}}{1-n_{3}} +\frac{n_{2}^{3}-3n_{2}{\bf n}_{2}\cdot{\bf n}_{2}}
{24\pi (1-n_{3})^{2}} \;.
\end{equation}
The density profile $\rho({\mathbf r})\equiv \rho(z)$, for planar walls, is obtained by solving the  Euler--Lagrange equation, i.e. from
\begin{equation}
\frac{\delta \Omega}{\delta\rho({\bf r})}=0\;.
\end{equation}
The solvation force ${\tilde f}_{solv}(L)$ (or excess pressure) can be obtained from
\begin{equation}
{\tilde f}_{solv}=-\frac{1}{A}\left(\frac{\partial\Omega_{ex}}{\partial L}\right)
\end{equation}
where $A$ denotes the area of the wall and $\Omega_{ex}\equiv \Omega+pV$ is the excess grand 
potential of the system. Here $p$ is the pressure of the reservoir at the chemical potential
$\mu$  and the  bulk density  $\rho_b$ and $V$ is the total volume.
Statistical mechanical sum rules \cite{Henderson92} for a confined fluid
lead to another expression for the solvation force
\begin{equation}
{\tilde f}_{solv}=-p-\int_{-\infty}^{\infty} dz \rho (z) \frac{\partial}{\partial z}V_{s}(z)
\end{equation}
We have used the above equation
to check the accuracy of the numerics.

Without loss of generality we chose the LJ $\sigma$ as a unit of length, and
introduce the following reduced units, $f_{solv}=\beta {\tilde f}_{solv}\sigma^{3}$,
$T^{\ast}=\frac{k_{B}T}{\varepsilon_{ff}}$, $\rho^{\ast}_b=\rho_b\sigma^{3}$.
For the system in question the reduced critical temperature and the
reduced critical density for the gas-liquid transition are
$T_c^{\ast}=1.319442$ and $\rho_c^{\ast}=0.245736$,
respectively.

We start by reporting the temperature dependence of the rescaled solvation
force $f_{solv}^{\ast}$, i.e. the  solvation force divided by its value at $T/T_c=0.615$,
for systems with $p=3,2$ and $L=50.4\sigma$ (see Fig.~\ref{fig:dft_t}).
In analogy with  the results presented in Sec.~\ref{subsec:DMRG}
the chemical potential $\mu(T)$ (or, equivalently the bulk density of the reservoir,
$\rho_{b}^{\ast}$) is fixed such that the bulk fluid  is slightly off 
 coexistence on the liquid side of the bulk coexistence curve.
We notice that for the system  
 with the wall-fluid interactions of the finite range, i.e. for truncated  wall-fluid potential
(\ref{eq:wp}) with $p=3$ and cut-off $z=2.5\sigma$,
(see the inset) the solvation force is extremaly  small away from $T_c$.
At low temperature  the system is on the oscillatory side of the Fisher-Widom line,
so  the solvation force should  decay  in an exponentially-damped
oscillatory fashion i.e.  $f_{solv}(L\to\infty)$ 
$\sim\exp(-a_0L)\cos(a_1 L)$~\cite{dijkstra:00:0}.
Consequently, for the large separations considered here the solvation force well below
$T_c$ becomes extremely small in magnitude and perishes in the numerical noise, 
i.e. its magnitude is  smaller than $10^{-10}$.
Close to the critical region a pronounced decrease 
in the solvation force is found (see the inset)
with its minimum located slightly above $T_c$. Note that for $T>T_c$ we follow the critical
isochore $\rho_b(\mu,\tau)=\rho_c$ and the decay is expected to be purely exponential, at least
in the range shown in Fig.~\ref{fig:dft_t}.
For this value of $L$ the magnitude is  extremely small.

For the systems
with long ranged wall-fluid potentials we find that for $p=3$ 
(Fig.~\ref{fig:dft_t}, solid line) the solvation force is positive away from
the critical region.
Only very close to $T_c$ does  $f_{solv}^{\ast}$ change sign and become  negative.
The minimum is located slightly above $T_c$. For still stronger  fluid-wall potentials 
i.e. for $p=2$ (Fig.~\ref{fig:dft_t}, dashed line) the solvation force is positive (repulsive)
 through  the entire range of temperatures under consideration.
The rescaled solvation force is almost identical
for both values of $p$ at temperatures away from  $T_c$ which is consistent with  the
result of  Eq.~(\ref{eq:LJasym}), i.e.  $f_{solv}^{\ast}\sim \rho(T)$ . As the systems
move towards the critical region, $f_{solv}^{\ast}$ begins to be different
for different powers of the wall-fluid potentials.
However the minimum of  $f_{solv}^{\ast}$ is located, similarly to previous cases,
slightly above the critical temperature. These findings
are in accordance with general considerations presented in the
Introduction.  The minimum of the solvation force is connected
with the critical Casimir effect \cite{krech:99:0}. For short ranged wall-fluid potentials
this effect is dominant (see the inset to Fig. ~\ref{fig:dft_t}) 
as the direct  influence of the wall is negligible at large separations.
When the long ranged wall-fluid
potentials are introduced, the contribution from the regular part
of the solvation force $f_{solv}^{reg}$ , which is different for different powers of the wall-fluid
potential,
dominates for temperatures away from the $T_c$ while for systems in the critical region
the singular part of the solvation force $f_{solv}^{sing}$ 
 comes into play. The dominant decay of the singular part is 
the same  for both value of  $p$. In the present DFT approach
the critical Casimir force is treated 
in  mean-field, therefore $f_{solv}^{sing}$  is given by (\ref{eq:fssCas}) with $d=4$ and the  mean-field
value of the exponent $\nu$ ($\nu_{MF}=1/2$).

In Figs.~\ref{fig:dft_hp3}-\ref{fig:dft_hp2} we show the $L$-dependence of the solvation force
for systems with long-range wall-fluid potentials for the bulk reservoir state  $T^{\ast}=1.0$,
$\rho_{b}^{\ast}=0.6148$. Again the temperature and chemical potential were fixed such
that the system is just slightly away from coexistence on the liquid branch of the
coexistence curve. We observe that for intermediate  $L$ the well-known oscillatory behaviour of $f_{solv}$
characteristic,  for short separations,  is  
damped and changes gradually to the power-law decay enforced by the external (long ranged)
wall-fluid potential. The asymptotic behaviour of the solvation force is demonstrated 
in Fig.~\ref{fig:dft_hp3}b and Fig.~\ref{fig:dft_hp2}b,
where we plot the logarithm of $f_{solv}$ as a function of
the logarithm of the inverse separation (symbols) along with  best straight line
 fits with imposed slope
(solid lines) for the systems with $p=3$, and $2$, respectively.
These are  similar to these in  Fig.~\ref{fig:r4},  presented
in Sec.~\ref{subsec:DMRG}. In both cases we find a nice agreement between the DFT results  and
the predictions  from Eq.~(\ref{eq:LJasym}). The asymptotic form of the
decay is quite  visible already for medium separations, i.e. for $L>50$.
In order to  investigate further the asymptotic
behaviour of the solvation force we have performed least-square fits assuming
a power-law, i.e. assuming $f_{solv}(L)=a_0 L^{\kappa}$ with $a_0, \kappa$
taken  as fit parameters.
For the system presented in Figs.~\ref{fig:dft_hp3}-\ref{fig:dft_hp2}
best fits performed for $190>L>110$ are $4.69x^{-2.99}$
and $4.62x^{-1.99}$, respectively, where $x=L/\sigma$. The value of the 
 constant $a_0$ predicted by Eq.~(\ref{eq:LJasym}) is
4.92 and  differs from those  given  by least-square fits by $\sim 6\%$.
We also note that the constant $a_0$  should be the same for both potentials
considered here (it depends on only the density $\rho_b^{\ast}$ and
the parameter $B=4\varepsilon f_w$ which is same for both) and indeed,
 to a good approximation, this is the case here.

It has been well established \cite{evans:86:0,evans:87:0} 
that if the chemical potential is fixed
such that the fluid is on the gas side of the liquid-gas  phase diagram
the  solvation force can (as a function
of the separation)  exhibit a jump that is a direct
manifestation of  capillary condensation. The discontinuity can appear for
both short ranged and long ranged wall-fluid potentials.
If the wall-fluid interactions are {\it short ranged}, $f_{solv}(L)$ will
change from small negative (weakly attractive) values at large $L$
(corresponding to the 'gas' phase) to larger negative values  for smaller $L$
(corresponding to the 'liquid' phase). As argued  in the Ref.~\cite{evans:87:0}
on the basis of macroscopic thermodynamic arguments there is always  the term $-\Delta\mu(\rho_l-\rho_g)$ in the expression for $f_{solv}$ for the capillary condensed 'fluid' which gives rise to the
aforementioned jump. On the basis of the theory of the
asymptotic decay of the correlation functions \cite{evans:93:0,evans:94:1}
it was argued  that the solvation force should
 decay (asymptotically) in the same manner as the bulk pair
correlation function. In the present case the fluid lies on the monotonic side of the Fisher-Widom
line.
A  schematic plot of  $f_{solv}$ is shown 
in Fig.~\ref{fig:dft_schem}a
(for clarity the oscillatory part of the solvation force, relevant for
small separations is omitted here).

When  the wall-fluid interactions become {\it long ranged} the solvation force should decay
asymptotically in the same fashion as the wall-fluid potential because its contribution
to the $f_{solv}$ will be dominant with respect to the fast-decaying (exponential)
fluid-fluid contribution. On the other hand, the discontinuity in $f_{solv}$
connected with capillary condensation must remain.
Thus one can anticipate that the solvation force should  behave in the manner
presented in Fig.~\ref{fig:dft_schem}b
(again, for clarity the oscillatory part of the solvation force,  relevant for
small separations is omitted here).
Namely, upon increasing the wall-wall separation $L$, the solvation force
jumps from larger to smaller but still negative values.
This discontinuity is associated with  capillary condensation.
Next, $f_{solv}$ changes  its sign (the point denoted by $Z$),
attains a maximum (the point denoted by $M$) followed by the inflection point
(denoted by $I$) and, finally, reaches the region of the asymptotic decay
dictated by Eq.~(\ref{eq:LJasym}).  The specific example of the solvation force
calculated by DFT for the system with $p=3$, $T^{\star}=1.2$ and $\rho^{\star}_{b}=0.06$
shown in Fig.~\ref{fig:dft_hp3cap} fits  well into the general behaviour described above.
The jump associated with the capillary condensation occuring
at $L\sim10.5$ (c.f. Fig.~\ref{fig:dft_hp3cap}a) is from larger to smaller but still negative
values of $f_{solv}$. As the wall-wall separation is increased 
(c.f. Fig.~\ref{fig:dft_hp3cap}b) $f_{solv}$
changes its sign around $L\sim72$, attains a maximum at $L\sim95$.
After an inflection point at $L\sim130$ the solvation force changes  slowly
towards its asymptotic form of decay. It must be mentioned however
that even for the largest separation studied ($L=190$) the asymptotics
(i.e. the decay with a power-law the same as the decay of the wall-fluid
potential) could not be still reached. We think  one needs to go to  the
separations as large as several hundreds to rich the asymptotic range.

\section{Summary and Conclusions}
\label{sec:sum}
In this paper we have performed calculations of  the solvation force for  
 an Ising film subject to  long ranged boundary fields 
(\ref{eq:bf}) with $p=50, 4, 3$, and 2, 
 and for  a truncated  LJ  fluid confined between
 two planar walls that exert a
$ 9-p$ with $ p=2, 3$   wall-fluid potential (\ref{eq:wp}).
For an Ising film results have been obtained in $d=2$ 
for  states along the 
line of the bulk two-phase coexistence $h=0$ by means of 
 the  DMRG method
which takes into account fluctuations of the order parameter.
 Since in two dimensions critical fluctuations  are particularly strong,
 results obtained in this model  serve as an ultimate test of the effects
 of fluctuations  on the behaviour  of the solvation force. 
For LJ fluid  results have been  obtained for several states  on a liquid
side  of the bulk two-phase coexistence and on the
critical isochore for $T>T_c$ within a nonlocal DFT which is
a  mean field theory. This approach  accounts for  packing effects 
and hence  oscillations in the solvation force.  

We observe  major differences in the behaviour of the solvation force between the
both models. In the Ising system $f_{solv}$ at low  temperatures
is positive (repulsive) and decays for large
$L$ in the same fashion as the boundary field, i.e., $f_{solv}=L^{-p}$,
whereas at high temperatures $f_{solv}$ is  negative (attractive)
and the asymptotic decay is of the higher order than that
 of the boundary field,
i.e.,  $f_{solv}=L^{-(p+1)}$.
 In the LJ fluid system  $f_{solv}$
is always repulsive away from the critical region
and decays for large $L$ with the the same power law 
as the wall-fluid potential, which is consistent with
 the general result~(\ref{eq:LJasym})
based on the wall-particle Ornstein-Zernike equations~\cite{attard:91:0}.
As discussed within a Landau approach in the Sec.~\ref{sec:landau} 
 the origin of this discrepancy is due to the specific symmetry
 of the Ising model with the spontaneous magnetization  $m^{\ast}$ equal to zero  above $T_c$; note that for  a fluid  $\rho _c(T)\ne 0$  above $T_c$.

Our results imply  that  for a confined  fluid 
with short ranged fluid-fluid interactions and long ranged wall-fluid potential
 the solvation force,  for large wall  separations,
 can be expressed  as a sum of a regular part $ f_{solv}^{reg}$, 
which decays in the same fashion as the 
wall-fluid potential, and a  part $ f_{solv}^{sing}$ 
 arising  from L-dependent singular
 contribution~\cite{krech:99:0}
to the free energy, which  is a close analog of the Casimir force in electromagnetism.
 The singular contribution to the free energy
is   responsible for the critical singularities  at the bulk critical point.
Thus, for ordering field $h=0$ and $L\to \infty$
\begin{equation}
\label{eq:dec}
f_{solv}\sim  f_{solv}^{reg}+f_{solv}^{sing}=2\rho  BL^{-p}+(d-1)L^{-d}W_{aa}(\tau L^{1/\nu})
\end{equation}
where $W_{aa}(y)$ is a finite-size scaling function for system with two identical walls $a$,
and $B=4\varepsilon_{fw}$ for a wall-fluid potential of a form (\ref{eq:wp}).
In mean-field  $d$ takes the value 4 in the  final term of Eq.~(\ref{eq:dec}).
Such a  decomposition applies also for an  Ising spins  system below $T_c$ 
with $\rho B$ replaced by $m^{\ast}h_1$.
Above $T_c$,  $m^{\ast}$ vanishes and the leading regular part of the solvation force
becomes $\sim L^{-(p+1)}$ as already mentioned above. 
The scaling function  $W_{aa}(y)$ is negative and vanishingly small away from the critical region, therefore, in $d=2$ and 3 and for all values of the parameter $p$ the positive  regular part of the solvation force dominates away
 from $T_c$.
Fluctuation induced attractive  Casimir force is particularly 
 strong  in  two dimensions
and dominates the behaviour of the solvation force in the critical region
for all considered values of the parameter $p$ - see Fig.~\ref{fig:fsT}. 
In mean field $f_{solv}^{sing}$ is much weaker and only when 
 $p\ge d=3$  the solvation force becomes weakly negative near  $T_c$. For $p=2$
the regular part $f_{solv}^{reg}$ gives the leading decay  and 
although the solvation force
exhibits minimum near $T_c$, it remains positive for all temperatures
 as is seen in Fig.~\ref{fig:dft_t}.

As a last remark we note that the inclusion of power-law
 fluid-fluid interactions would   modify  our results. This
 can be infered from
the asymptotic  integral equation results for the solvation force of the full
LJ fluid - see Eq.~(\ref{eq:attard}). The final term is associated with the $-r^{-n}$
decay of the fluid-fluid potential. For LJ $n=6$ so Eq.~(\ref{eq:attard})   predicts 
 an additional 
 {\it attractive } term   $\sim L^{-3}$ in the solvation force. This should
then be included  in  Eq.~(\ref{eq:dec}).
This contribution  competes with other terms and may lead to even  richer
behaviour of the solvation force.  
We want to stress that in our studies we have neglected the contribution due to the direct wall-wall
interaction potential~\cite{footnote}.

After the completion of our calculations  we  learnt of a recent
  article by Pertsin and Grunze~\cite{pertsin:03:0} 
describing  the results of Monte Carlo simulations  of a
Lennard-Jones fluid confined between two planar walls. Since their results were at complete variance with ours and with the general predictions of Eq.~(\ref{eq:attard}) we decided to perform DFT calculations for the same state point  as in the simulations of Ref.~\cite{pertsin:03:0}. The results  are presented in Appendix B. As expected, we find no evidence  for the extremely long ranged solvation
force reported in Ref.~\cite{pertsin:03:0}. We have since ascertained, via correspondence  with 
Professor A. Pertsin and Professor R. Evans, that the simulations reported in  Ref.~\cite{pertsin:03:0}
were flawed and that the authors shall publish an erratum pointing this out.
The new results, however,  are in a good qualitative agreement with our  DFT
results.

\begin{acknowledgments}
This work was partially funded by KBN grant Nos. 4T09A06622 and 2P03B10616.
We have benefitted from conversations with R. Evans, A. Ciach, 
J. Sznajd and  D. Dantchev.
Comments of A. O. Parry prompted us to analyse the Landau model.
We wish  to thank  R. Evans for the critical reading of the  manuscript.
\end{acknowledgments}

\appendix*
\section{A}

In order to  evaluate contributions to the solvation force arising from 
  terms (\ref{eq:lan103}), (\ref{eq:lan11}) and (\ref{eq:lan111}) we use 
Eqs.~(\ref{eq:solfordef}),~(\ref{eq:excess}), and ~(\ref{eq:lan1}). 

First we integrate over $z$ to obtain   contributions to  the total 
free energy. We find~\cite{grad}
\begin{equation}
\label{eq:a0}
\int_0^L\frac{dz}{(\lambda+z)^p(L+\lambda-z)^p}=\frac{L}{2p-1}\sum_{k=0}^{p-2}\frac{2^{k+1}(2p-1)(2p-3)\cdots(2p-1-2k)}{(p-1)(p-2)\cdots (p-k-1)\Delta^{k+1}(L\lambda+\lambda)^{p-k-1}},
\end{equation}
where $\Delta=(4L\lambda+4\lambda+L^2)$.
In the limit of $L\to\infty$ and $\lambda/L\to 0$  the above integral simplifies
\begin{equation}
\label{eq:a01}
\int_0^L\frac{dz}{(\lambda+z)^p(L+\lambda-z)^p}\to_{L\to\infty}\sum_{k=0}^{p-2}\frac{2^{k+1}(2p-3)\cdots(2p-1-2k)}{(p-1)(p-2)\cdots (p-k-1)\lambda^{p-k-1}L^{p+k}}.
\end{equation}
It follows that the contribution  to the solvation force arising from 
the   term  (\ref{eq:lan103}) is 
\begin{equation}
\label{eq:a02}
\frac{\xi_b^2h_1^2}{b}\frac{\partial}{\partial L}\int_0^Ldz\frac{1}{(\lambda+z)^{p}(L+\lambda-z)^p}=-\frac{2\xi_b^2h_1^2}{b\lambda^{p-1}}\frac{p}{p-1}\frac{1}{L^{p+1}}+O(L^{-(p+2)}),~~~L\to\infty.
\end{equation}

Integration of terms  (\ref{eq:lan11}) yields
\begin{equation}
\label{eq:a1}
\int_0^Ldz\frac{e^{-z/\xi_b}}{(L+\lambda-z)^{p}}= \int_0^Ldz\frac{e^{-(L-z)/\xi_b}}{(\lambda-z)^{p}},
\end{equation}
and 
\begin{eqnarray}
\label{eq:a2}
\int_0^Ldz\frac{e^{-z/\xi_b}}{(L+\lambda-z)^{p}}&=&\frac{1}{\xi_b^{p-1}(p-1)!}\sum_{k=1}^{p-1}\xi_b^k(k-1)!\left[\frac{e^{-L/\xi_b}}{\lambda^k}-\frac{1}{(L+\lambda)^k}\right]  \nonumber \\
&+&\frac{1}{\xi_b^{p-1}(p-1)!}e^{-(L+\lambda)/\xi_b}\left[Ei(\frac{L+\lambda}{\xi_b})-Ei(\frac{\lambda}{\xi_b})\right],
\end{eqnarray}
where $Ei(x)$ is the Exponential-Integral Function.

An asymptotic representation of the function  $Ei(x)$
 for large $x$ is~\cite{grad}
\begin{equation}
\label{eq:a3}
e^{-x}Ei(x)=\sum_{k=1}^n\frac{(k-1)!}{x^k}+O(x^{-(n+1)})~~~~(x\to\infty).
\end{equation}
Thus,  in a limit of  large $L$, the second term in  the expression for the
integral (\ref{eq:a2}) behaves as 
\begin{eqnarray}
\label{eq:a4}
\frac{1}{\xi_b^{p-1}(p-1)!}e^{-(L+\lambda)/\xi_b}\left[Ei(\frac{L+\lambda}{\xi_b})-Ei(\frac{\lambda}{\xi_b})\right]&=& \nonumber \\ 
\frac{1}{\xi_b^{p-1}(p-1)!}\left[\sum_{k=1}^{n}\xi_b^k(k-1)!\frac{1}{(L+\lambda)^k}-Ei(\frac{\lambda}{\xi_b})\right]&+&O((L+\lambda)^{-(n+1)})~~~(L\to\infty).
\end{eqnarray}
Using the above asymptotic expression we obtain the asymptotic 
behaviour  of the integral (\ref{eq:a2}) in a limit of $L\to \infty$
\begin{eqnarray}
\label{eq:a5}
\int_0^Ldz \frac{e^{-z/\xi_b}}{(L+\lambda-z)^{p}} &=&\frac{e^{-L/\xi_b}}{\xi_b^{p-1}(p-1)!}\left[ \sum_{k=1}^{p-1}\xi_b^k(k-1)!\frac{1}{\lambda^k}- e^{-\lambda/\xi_b}Ei(\frac{\lambda}{\xi_b})\right] \nonumber \\ &+&\frac{\xi_b}{(L+\lambda)^{p}}+O(L^{-(p+1)}).
\end{eqnarray}
It follows that 
\begin{eqnarray}
\label{eq:a6}
-\frac{\partial}{\partial L}\int_0^Ldz \frac{e^{-z/\xi_b}}{(L+\lambda-z)^{p}}&=&\frac{e^{-L/\xi_b}}{\xi_b^{p}(p-1)!}\left[ \sum_{k=1}^{p-1}\xi_b^k(k-1)!\frac{1}{\lambda^k}- e^{-\lambda/\xi_b}Ei(\frac{\lambda}{\xi_b})\right] \nonumber \\&+&\frac{p\xi_b}{(L+\lambda)^{p+1}}+O(L^{-(p+2)})~~~~(L\to \infty).
\end{eqnarray}
Now, consider contributions to the solvation force arising from 
  terms
\begin{equation}
\label{eq:a7}
\frac{ e^{-z/\xi_b}}{(\lambda+z)^{p}};~~~~~~~~~~~~~~~~\frac{ e^{-(L-z)/\xi_b}}{(L+\lambda-z)^{p}}.
\end{equation}
We have
\begin{equation}
\label{eq:a8}
\int_0^Ldz\frac{e^{-z/\xi_b}}{(\lambda+z)^{p}}=\int_0^Ldz\frac{e^{-(L-z)/\xi_b}}{(L+\lambda-z)^{p}}
\end{equation}
and
\begin{equation}
\label{eq:a9}
-\frac{\partial}{\partial L}\int_0^Ldz\frac{e^{-z/\xi_b}}{(\lambda+z)^{p}}=-\frac{e^{-L/\xi_b}}{(\lambda+L)^{p}}+\frac{1}{\lambda ^{p}}.
\end{equation}

\section{B}
In order to compare with the results of Ref.~\cite{pertsin:03:0} we carried out DFT calculations for 
$T^{\ast}=1.0$ and  $\rho^{\ast}_b=0.6853$, the simulation state point. In Ref.~\cite{pertsin:03:0}
the LJ potential is cut-off at $r_{cut}=2.5\sigma$, the same as in our DFT, and the wall-fluid potentials have the same form as (\ref{eq:wp}). Fig.~\ref{fig:appen}a shows our results for $f_{solv}(L)$ for the case of a 9-3 wall-fluid potential, i.e. $p=3$, whilst  Fig.~\ref{fig:appen}b
refers to the {\it truncated} 9-4 wall-potential, set to zero at $z=2.5\sigma$.
In the first case we confirmed that for large $L$, $f_{solv}(L)$ is positive and decays as $L^{-3}$.
Indeed the fit to this power law is as good as that for the neighbouring state point shown in Fig.~\ref{fig:dft_hp3}. For the truncated 9-4 potential $f_{solv}(L)$ exhibits the expected exponentially
damped oscillatory decay for large $L$. Thus our results for this state point are also in
agreement  with the general theory described  
 in the Introduction. By contrast Pertsin and Grunze~\cite{pertsin:03:0} find for large $L$ an 
{\it attractive} solvation force which decays very slowly, roughly as $L^{-1}$, for both 
the 9-3 and the truncated 10-4 wall-fluid potentials (see their Fig. 3a). Such behaviour
would seem to be unphysical.

\begin{figure}
\includegraphics[width=16cm,clip]{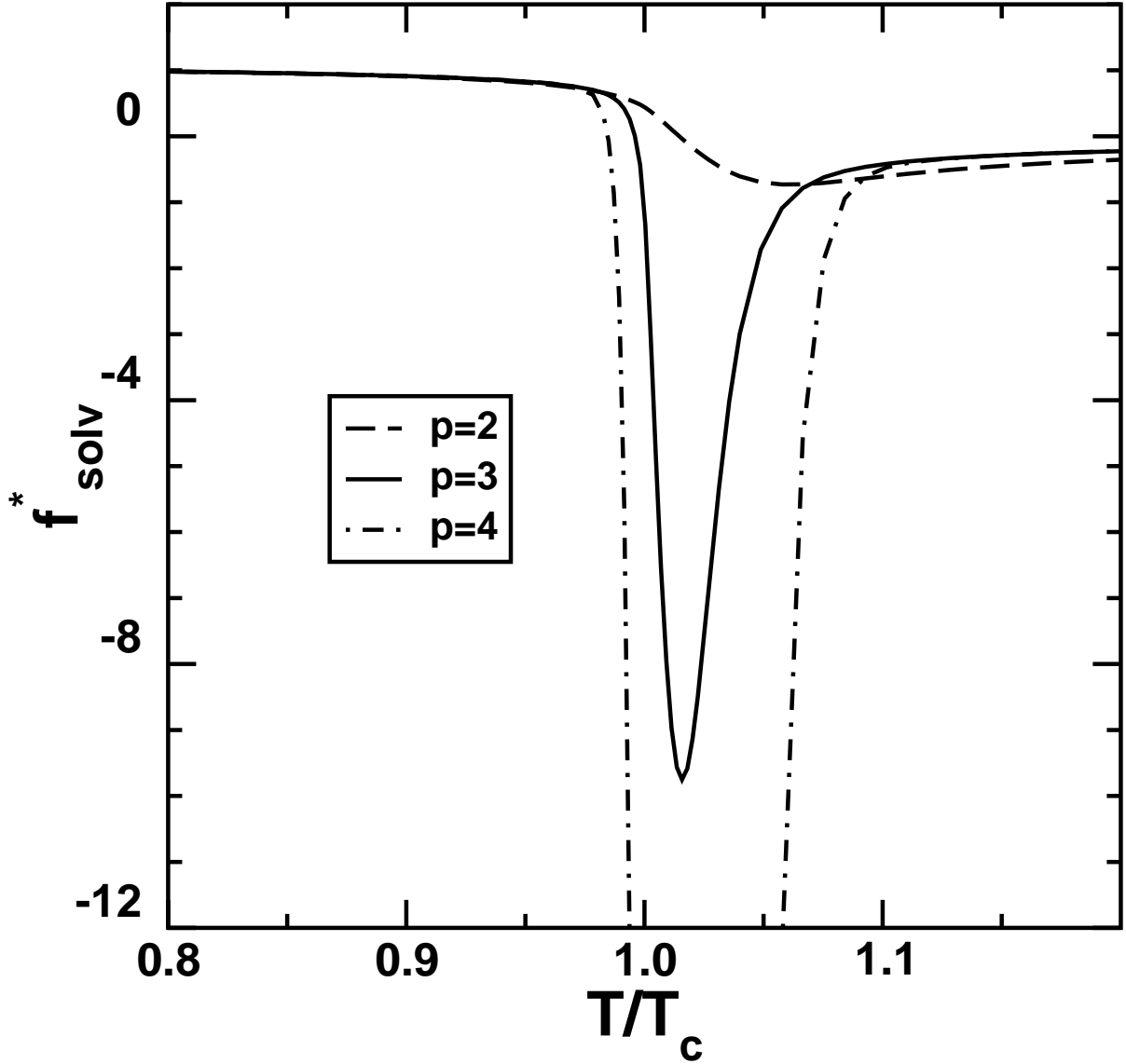}
\caption{\label{fig:r4}
Solvation force  for the Ising strip of width $L=100$, $h_1\approx 8.16$
rescaled with  its value at $T/T_c=0.75$, plotted  as a function of the reduced temperature.
Below $T_c$, the results   calculated  for  three  different values
of the power $p$ lie on a  common curve $\sim m^*(T/T_c)$, where $m^*(T)$ 
is bulk spontaneous magnetization. This behaviour  agrees with the prediction
 Eq.~(\ref{eq:LJasym}) 
with $2\rho B$ replaced by $2m^*(T)h_1$. 
}
\end{figure}

\begin{figure}
\includegraphics[width=16cm,clip]{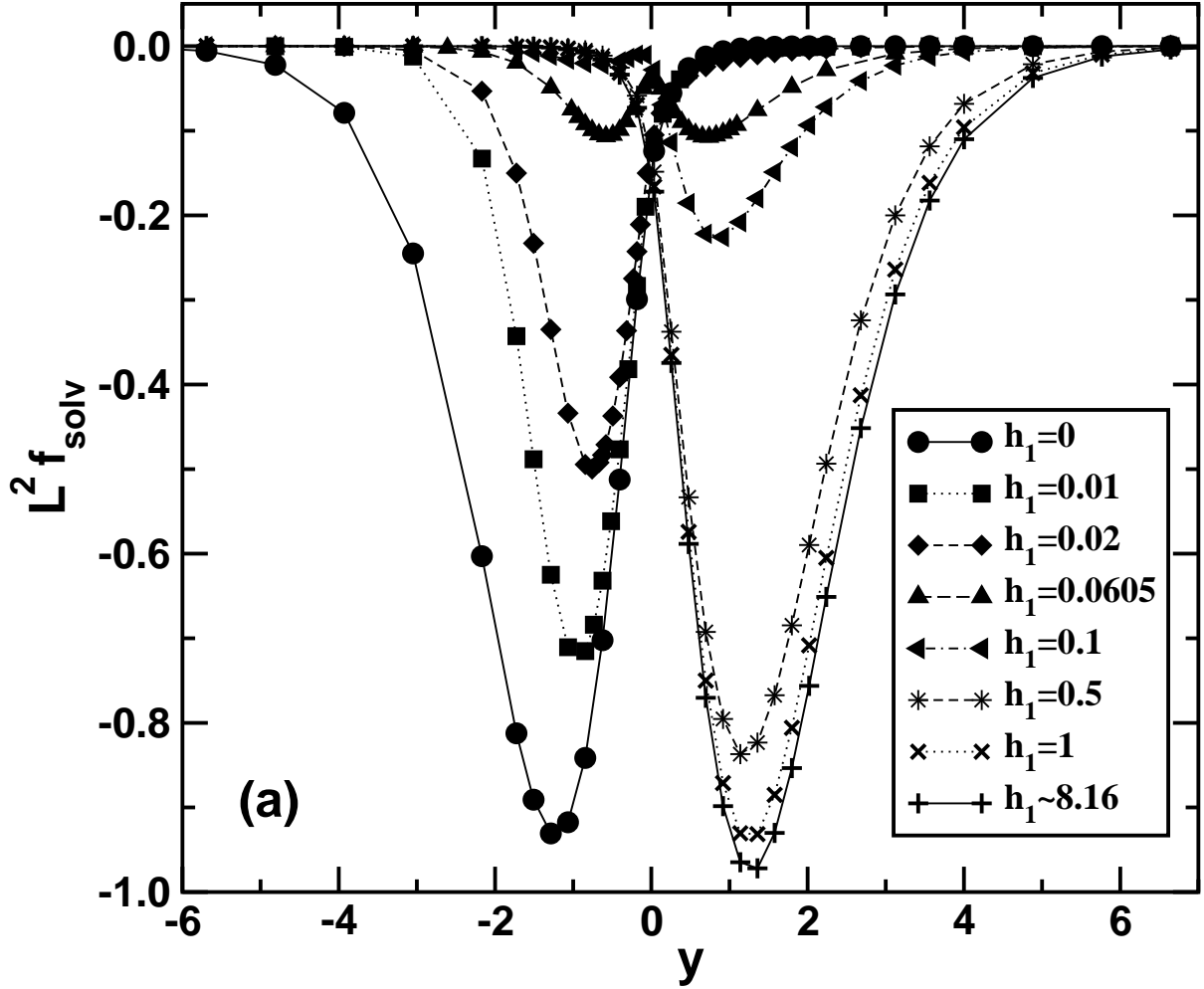}\\
\caption{\label{fig:fsTh_1}
Solvation force (in units of $J$)
for the same system as in Fig.~\ref{fig:r4} as a function
of the variable $y$  for a  selection of values of  the strength
parameter $h_1$  and  for two values of the  parameter $p$:
({\bf a}) $p=50$ and  ({\bf b}) $p=2$. Results
for different values of $h_1$  in Fig.~\ref{fig:fsTh_1}b are
 represented by the same symbols as in  Fig.~\ref{fig:fsTh_1}a.
 It is seen that the symmetry, Eq.~(\ref{eq:scpr})  between $f_{solv}(y)$ for the
 free ($h_1=0$) and $(++)$  ($h_1\sim 8.16$) boundary conditions
 breaks when the boundary fields become long ranged.  For weak  $h_1$
 the solvation force exhibit two minima.
}
\end{figure}

\newpage
\begin{center}
{\large Figure Captions}
\end{center}
\begin{itemize}
\item{Fig.1} Solvation force (in units of $J$) for
 an
 Ising strip of the  width $L=300$ at vanishing bulk magnetic field  subject to
 long ranged  boundary  fields $H_l=H^s_l+H^s_{L+1-l}$, $H^s_l=h_1/l^p$
 as a function of the scaling variable $y\equiv \tau (L)^{1/\nu}=\tau L$
 for various values of  $p$
  and the amplitude $h_1\approx 8.16$ calculated using DMRG method.
  The inset shows, on an expanded scale, that results for $p=50$ and $p=4$
  differ only very slightly in the near-critical region.
   For all values of $p$ $~f_{solv} $ is attractive above $T_c$ and, except
   for $p=50$,  repulsive
   below $T_c$.

\item{Fig.2}
Log-log plot of the solvation force (in units of $J$)
versus $1/L$  for two values of the
power $p$ describing the  decay  of the boundary field $H^s_l$, $p=2$ and 4.
({\bf a}) displays DMRG results for $T=0.79 T_c$ (filled symbols)  and $T=T_c$
(unfilled  symbols).  Straight lines correspond to $f_{solv}(L)\sim  L^{-p}$.
Data at $T_c$ align parallel to the line of a slope 2 showing the result
 $f_{solv}(L)\sim  L^{-d}$ (see Eq.~(1)) with $d=2$.
({\bf b}) displays DMRG results for $T=1.23T_c$. Straight lines correspond to $f_{solv}(L)\sim  L^{-(p+1)}$.

\item{Fig.3}
Solvation force  for the Ising strip of width $L=100$, $h_1\approx 8.16$
rescaled with  its value at $T/T_c=0.75$, plotted  as a function of the reduced temperature.
Below $T_c$, the results   calculated  for  three  different values
of the power $p$ lie on a  common curve $\sim m^*(T/T_c)$, where $m^*(T)$ 
is bulk spontaneous magnetization. This behaviour  agrees with the prediction
 Eq.~(\ref{eq:LJasym}) 
with $2\rho B$ replaced by $2m^*(T)h_1$.

\item{Fig.4} 
Solvation force (in units of $J$)
for the same system as in Fig.~\ref{fig:r4} as a function
of the variable $y$  for a  selection of values of  the strength
parameter $h_1$  and  for two values of the  parameter $p$:
({\bf a}) $p=50$ and  ({\bf b}) $p=2$. Results
for different values of $h_1$  in Fig.~\ref{fig:fsTh_1}b are
 represented by the same symbols as in  Fig.~\ref{fig:fsTh_1}a.
 It is seen that the symmetry, Eq.~(\ref{eq:scpr})  between $f_{solv}(y)$ for the
 free ($h_1=0$) and $(++)$  ($h_1\sim 8.16$) boundary conditions
 breaks when the boundary fields become long ranged.  For weak  $h_1$
 the solvation force exhibit two minima.

\item{Fig.5}
({\bf a}) Specific heat $C_H$ (in units of $k_B$)
and ({\bf b}) the inverse longitudinal spin-spin correlation
length $\xi_{\parallel}^{-1}$ 
for the same system as in Fig.~\ref{fig:r4}
as a function of the scaling variable $y$  for $h_1\approx 8.16$
and various values of the parameter $p$ (represented by the same
symbols for both quantities).

\item{Fig.6}
Solvation force  rescaled with its value at $T/T_c=0.615$, 
$f_{solv}^{\ast}\equiv f_{solv}(T/T_c)/f_{solv}(0.615)$,
as a function of the  temperature for systems with
long ranged fluid-wall potentials for $L=50.4\sigma$ evaluated from
density functional theory.
The chemical potential $\mu (T)$ is chosen so that the reservoir density
  bulk density $\rho_{b}^{\ast}(T)$
is  close to that of the bulk coexisting liquid for $T<T_c$ and $\rho_b(\mu,T)=\rho_c$ 
for $T\ge T_c$.
Solid line corresponds to the system with $p=3$, while dashed line is for the system
with $p=2$. The inset shows the temperature dependence of $f_{solv}$ for the system with
fluid-wall interactions of the finite range.

\item{Fig.7}
The oscillatory (part a) and asymptotic shown on a $\log-\log$ plot (part b)
regions of the solvation force for the system with $p=3$ evaluated
from density functional theory. The temperature ($T^{\ast}=1.0$)
and the reservoir density ($\rho_{b}^{\ast}=0.6148$) are fixed such that the fluid
is just slightly off the coexistence on the liquid branch of the 
coexistence curve.

\item{Fig.8}
The oscillatory (part a) and asymptotic  shown on a $\log-\log$ plot (part b)
regions of the solvation force for the system with $p=2$
evaluated from density functional theory. The temperature ($T^{\ast}=1.0$)
and the reservoir density ($\rho_{b}^{\ast}=0.6148$)
are fixed such that the fluid is just slightly off the coexistence
on the liquid  branch of the 
coexistence curve. Note that for $p=2$ the solvation force has more pronounced oscillations
than for   $p=3$.

\item{Fig.9}
Schematic plot of the solvation force
for  systems with short ranged  (part a) and long ranged (part b)
fluid-wall  interactions in the presence of  capillary condensation.
Dots denoted by $Z,M$ and $I$ in part b mark characteristic points of the solvation
force: the zero, the maximum and the inflection point, respectively.

\item{Fig.10}
The short distance  (part a) and asymptotic (part b)
regions of the solvation force for the system with long ranged  ($p=3$)
wall-fluid  interactions. Note the change of scale in each part.
 Capillary condensation gives rise to the jump at $L=10.5$.
For $L<10.5$ the confined fluid  is a 'liquid' which can exhibit oscillations. The bulk reservoir corresponds to  ($T^{\ast}=1.2$,
$\rho_{b}^{\ast}=0.06$).

\item{Fig.11}
Solvation force for the LJ fluid with the short ranged and long ranged
wall-fluid potentials calculated within DFT for $T^{\ast}=1.0$
and  $\rho^{\ast}_b=0.6853$, the same  state point as in 
the simulations by Pertsin and Grunze~\cite{pertsin:03:0}.
({\bf a}) is for the full 9-3 LJ potential.
Oscillations characteristic for short wall-wall separations
are gradually damped  and changed  into the power law asymptotic decay.
({\bf b}) is for the truncated  9-4 LJ potential with the cut-off
distance 2.5$\sigma$. Oscillations around 0 are present 
for all wall-wall separations.

\end{itemize}

\newpage

\begin{figure}
\includegraphics[width=16cm,clip]{r2d.eps}
\end{figure}

\begin{figure}
\includegraphics[width=16cm,clip]{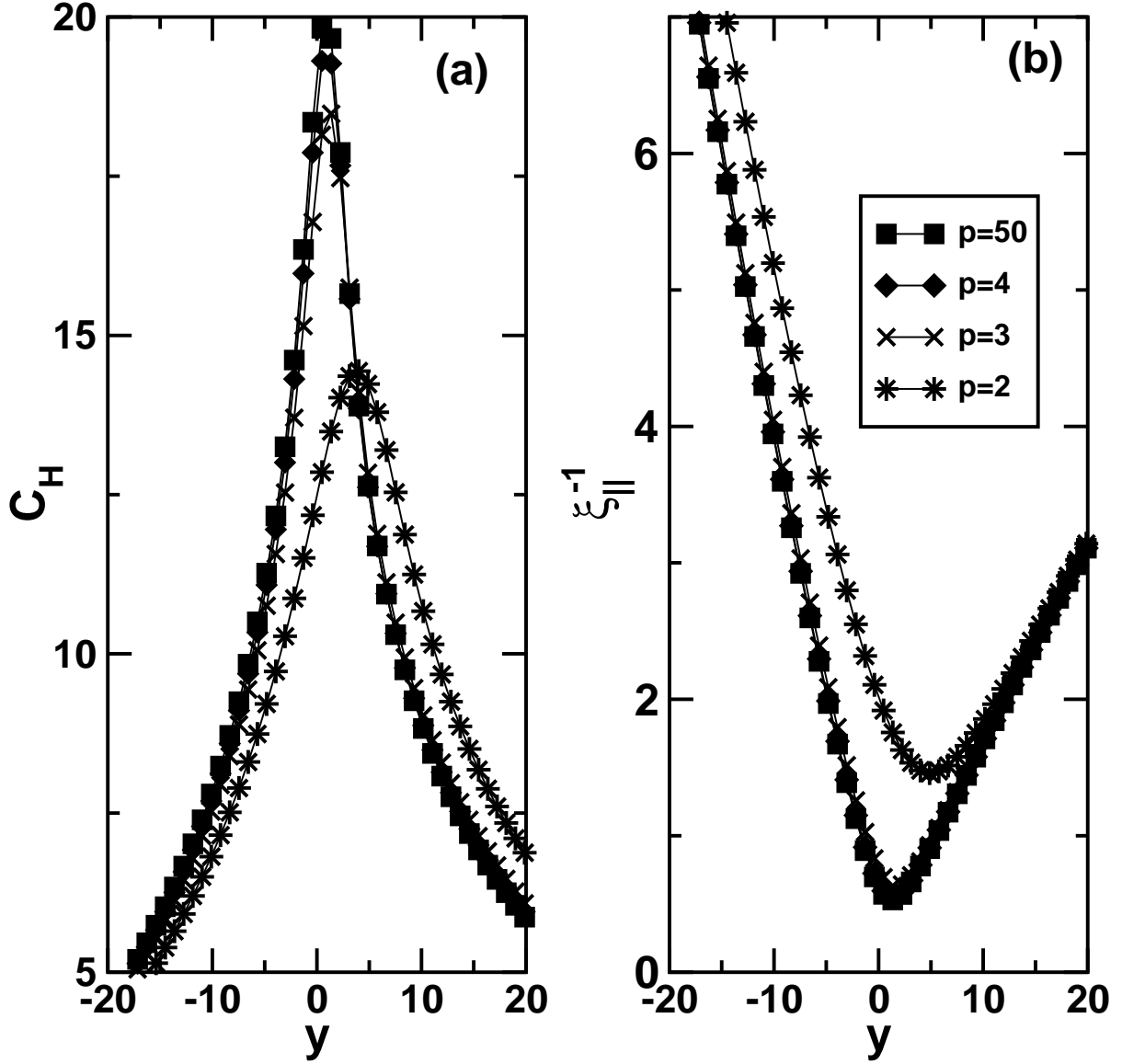}
\caption{\label{fig:rysCvXi} ({\bf a}) Specific heat $C_H$ (in units of $k_B$)
and ({\bf b}) the inverse longitudinal spin-spin correlation
length $\xi_{\parallel}^{-1}$ 
for the same system as in Fig.~\ref{fig:r4}
as a function of the scaling variable $y$  for $h_1\approx 8.16$
and various values of the parameter $p$ (represented by the same
symbols for both quantities).
}
\end{figure}

\begin{figure}
\includegraphics[width=16cm,clip]{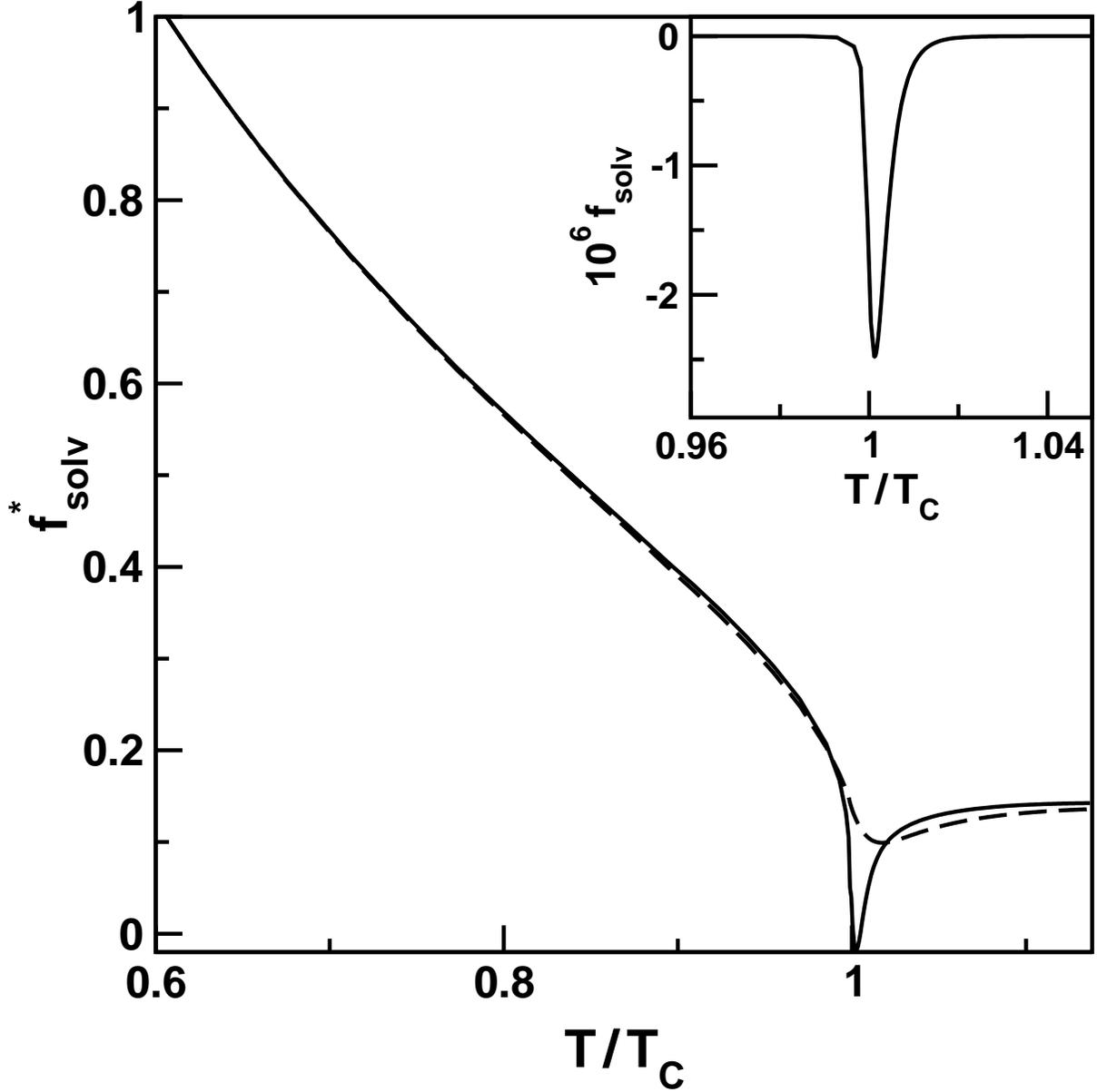}
\caption{\label{fig:dft_t} Solvation force  rescaled with its value at $T/T_c=0.615$, 
$f_{solv}^{\ast}\equiv f_{solv}(T/T_c)/f_{solv}(0.615)$,
as a function of the  temperature for systems with
long ranged fluid-wall potentials for $L=50.4\sigma$ evaluated from
density functional theory.
The chemical potential $\mu (T)$ is chosen so that the reservoir density
  bulk density $\rho_{b}^{\ast}(T)$
is  close to that of the bulk coexisting liquid for $T<T_c$ and $\rho_b(\mu,T)=\rho_c$ 
for $T\ge T_c$.
Solid line corresponds to the system with $p=3$, while dashed line is for the system
with $p=2$. The inset shows the temperature dependence of $f_{solv}$ for the system with
fluid-wall interactions of the finite range.
}
\end{figure}

\begin{figure}
\includegraphics[width=16cm,clip]{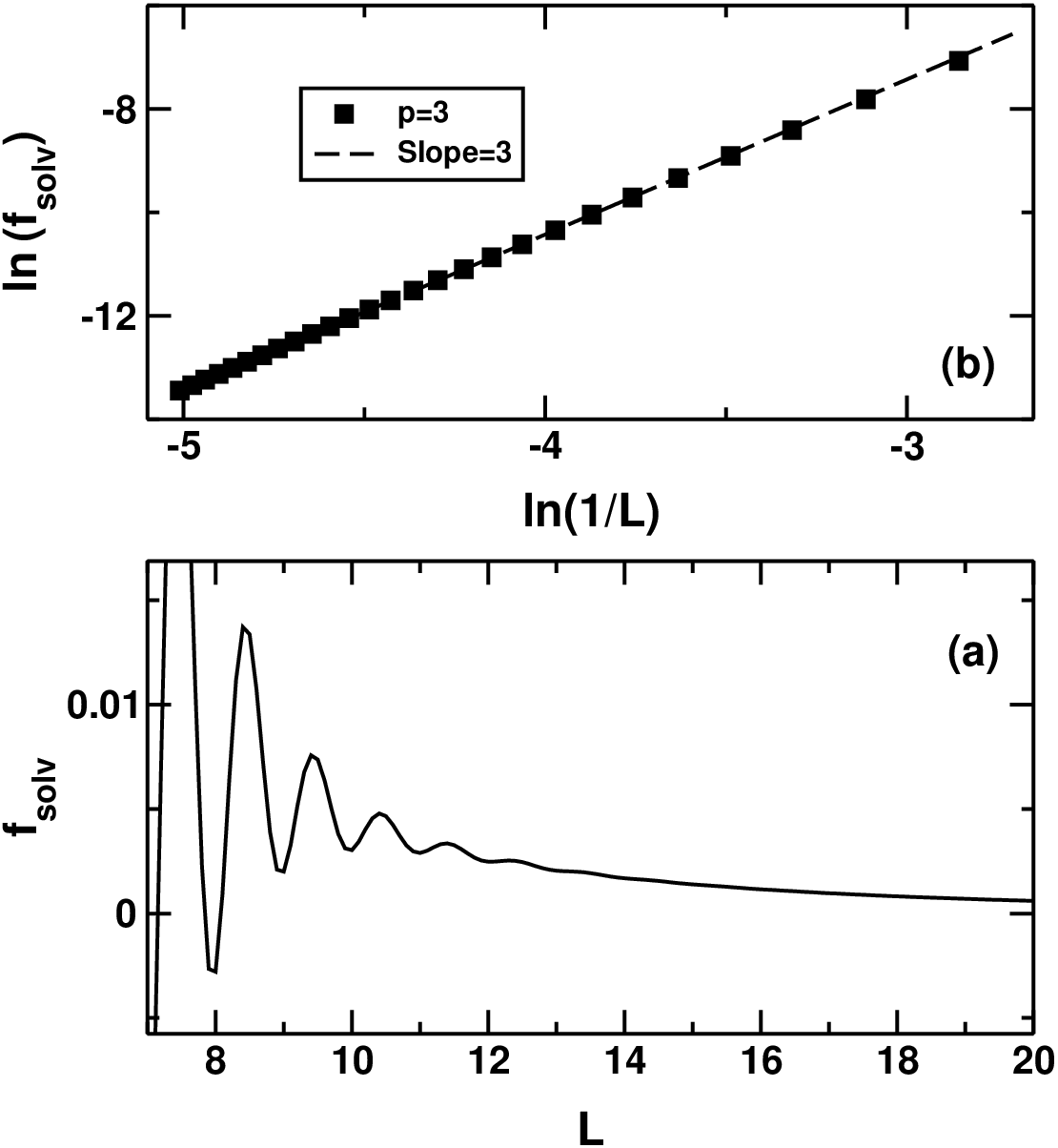}
\caption{\label{fig:dft_hp3} The oscillatory (part a) and asymptotic shown on a $\log-\log$ plot (part b)
regions of the solvation force for the system with $p=3$ evaluated
from density functional theory. The temperature ($T^{\ast}=1.0$)
and the reservoir density ($\rho_{b}^{\ast}=0.6148$) are fixed such that the fluid
is just slightly off the coexistence on the liquid branch of the 
coexistence curve.
}
\end{figure}

\begin{figure}
\includegraphics[width=16cm,clip]{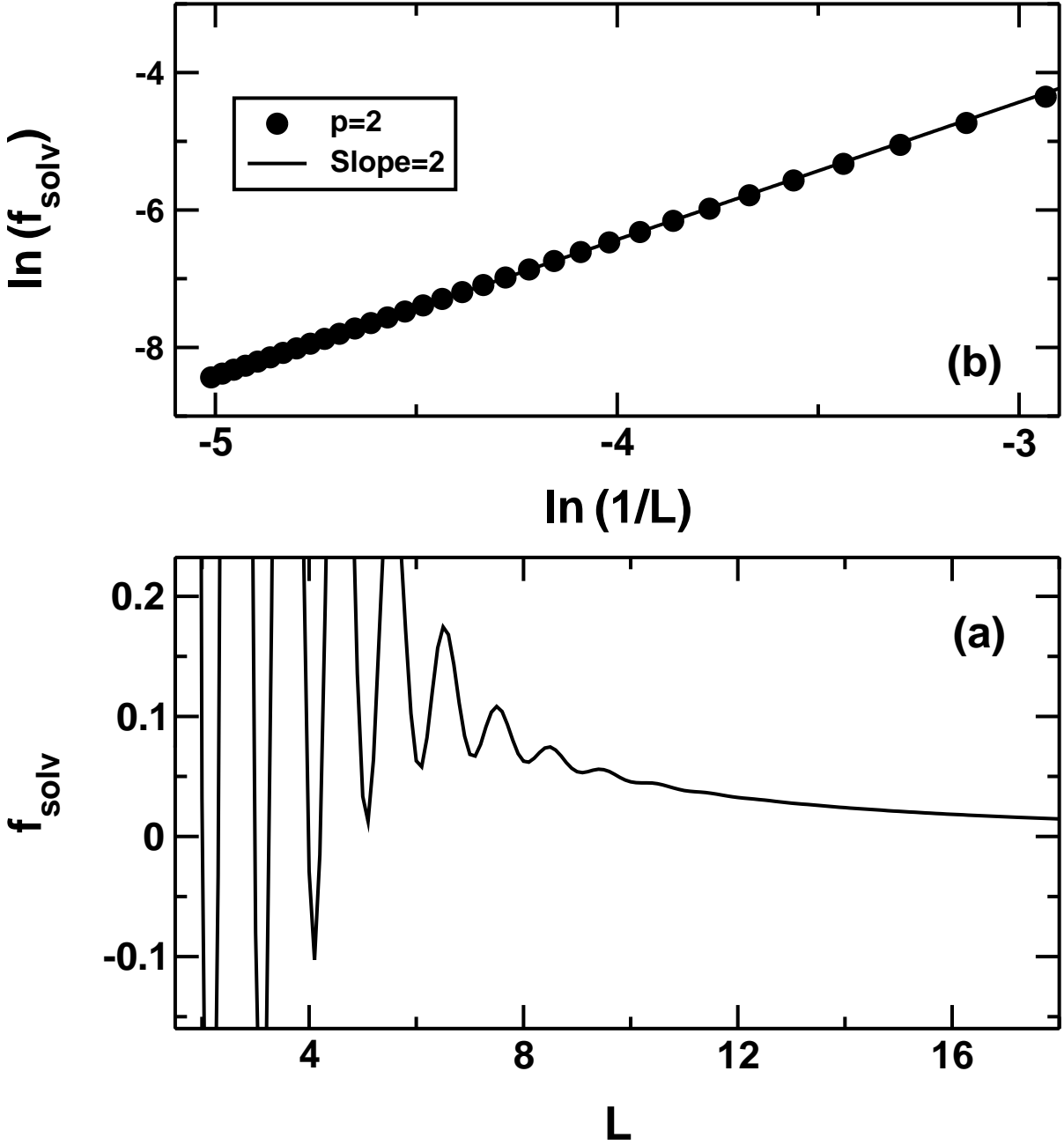}
\caption{\label{fig:dft_hp2}
The oscillatory (part a) and asymptotic  shown on a $\log-\log$ plot (part b)
regions of the solvation force for the system with $p=2$
evaluated from density functional theory. The temperature ($T^{\ast}=1.0$)
and the reservoir density ($\rho_{b}^{\ast}=0.6148$)
are fixed such that the fluid is just slightly off the coexistence
on the liquid  branch of the 
coexistence curve. Note that for $p=2$ the solvation force has more pronounced oscillations
than for $p=3$.
}
\end{figure}

\begin{figure}
\includegraphics[width=10cm,clip]{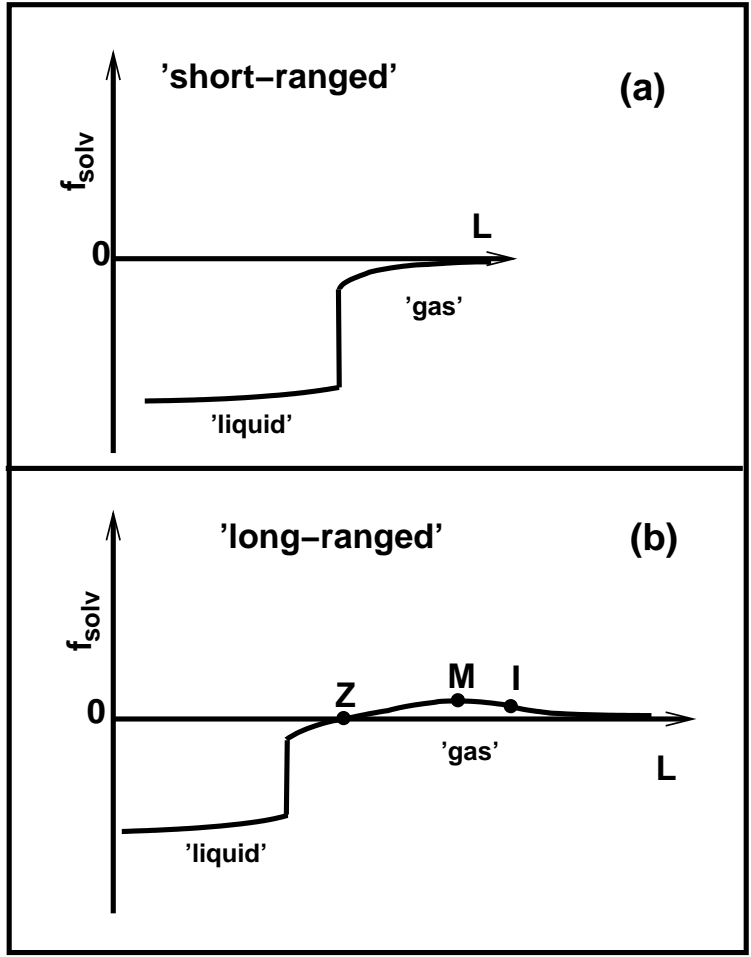}
\caption{\label{fig:dft_schem} 
Schematic plot of the solvation force
for  systems with short ranged  (part a) and long ranged (part b)
fluid-wall  interactions in the presence of  capillary condensation.
Dots denoted by $Z,M$ and $I$ in part b mark characteristic points of the solvation
force: the zero, the maximum and the inflection point, respectively.
}
\end{figure}

\begin{figure}
\includegraphics[width=16cm,clip]{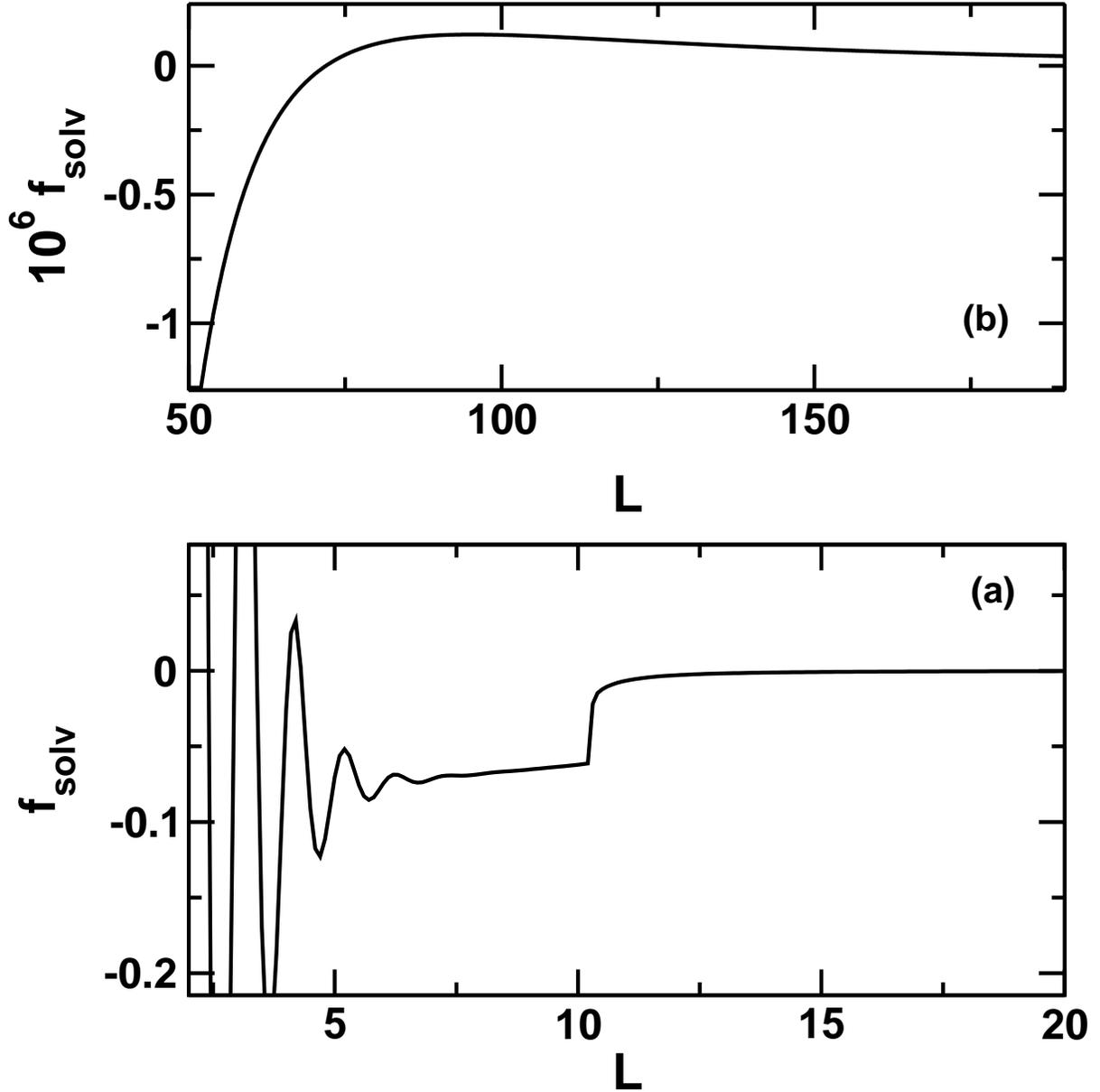}
\caption{\label{fig:dft_hp3cap}
The short distance  (part a) and asymptotic (part b)
regions of the solvation force for the system with long ranged  ($p=3$)
wall-fluid  interactions. Note the change of scale in each part.
 Capillary condensation gives rise to the jump at $L=10.5$.
For $L<10.5$ the confined fluid  is a 'liquid' which can exhibit oscillations. The bulk reservoir corresponds to  ($T^{\ast}=1.2$,
$\rho_{b}^{\ast}=0.06$).
}
\end{figure}

\begin{figure}
\includegraphics[width=16cm,clip]{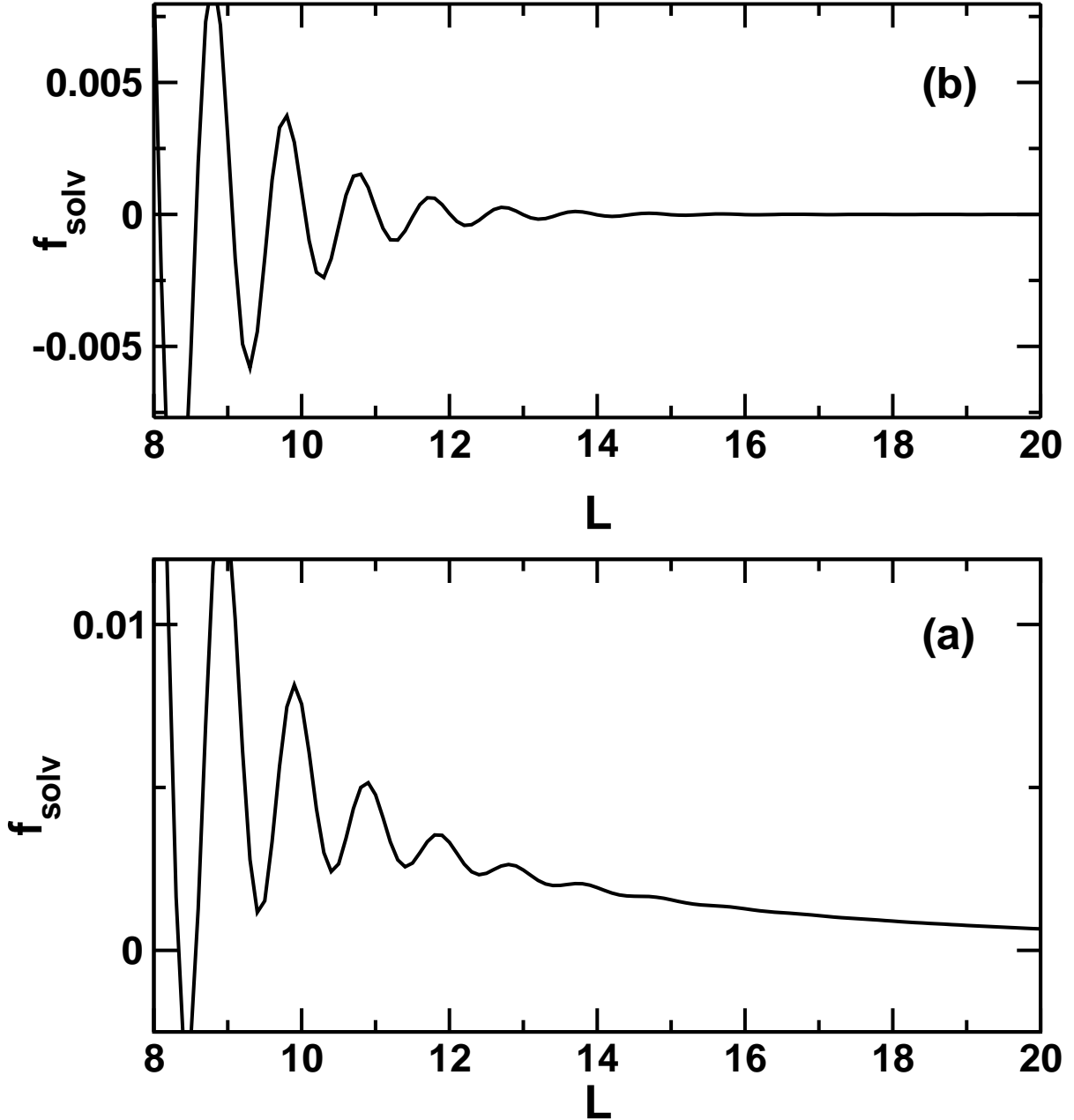}
\caption{\label{fig:appen} Solvation force for the LJ fluid with the short ranged and long ranged wall-fluid potentials calculated within DFT for $T^{\ast}=1.0$ and  $\rho^{\ast}_b=0.6853$,
 the same  state point as in  the simulations by Pertsin and Grunze~\cite{pertsin:03:0}.
({\bf a}) is for the full 9-3 LJ potential. Oscillations characteristic  for short wall-wall separations
are gradually damped  and changed  into the power law asymptotic decay.
({\bf b}) is for the truncated  9-4 LJ potential with the cut-off distance 2.5$\sigma$. 
Oscillations around 0 are present for all wall-wall separations.
}
\end{figure}


\begin{thebibliography}{99}

\bibitem{evans:90:0} R. Evans,  J.Phys.:Condens.Matter. {\bf 2}, 8989 (1990)
and references therein.

\bibitem{israelachvili:91:0} J.N Israelachvili, {\it Intermolecular and Surface
Forces} (Academic, London, 1991), 2nd ed.

\bibitem{israelachvili:78:0} J.N Israelachvili and G.E. Adams, JCS Faraday Trans. I {\bf 74}, 975 (1978);
 J.N Israelachvili, Accounts Chem. Res. {\bf 20}, 415 (1987).

\bibitem{smith:88:0} C.P. Smith, M. Maeda, L. Atanasoska, H.S. White, and D.J. McClure, J. Phys. Chem. {\bf 92}, 199 (1988).

\bibitem{parker:88:0} J.L. Parker and H.K. Christenson, J. Chem. Phys. {\bf 88}, 8013 (1988).

\bibitem{lee:89:0} C.S. Lee and G. Belfort, Proc. Natl. Acad. Sci USA {\bf 86}, 8392 (1989).

\bibitem{evans:90:1} See, e.g., R. Evans, in {\it Liquids at Interfaces}, 
Les Houches Session XLVIII, edited by J. Charvolin, J. Joanny, and J. 
Zinn-Justin (Elsevier, Amsterdam, 1990), p. 3.

\bibitem{evans:92:0} R. Evans, D.C. Hoyle, and A.O. Parry, Phys. Rev. A {\bf 45}, 3823 (1992).

\bibitem{henderson:92:0} J.R. Henderson and Z.A. Sabeur, J. Chem. Phys. {\bf 97}, 6750 (1992).

\bibitem{evans:93:0} R. Evans, J.R. Henderson, D.C. Hoyle, A.O. Parry, and Z.A. Sabeur, Molec. Phys. {\bf 80}, 755 (1993).

\bibitem{evans:94:1} R. Evans, R.J.F. Leote de Carvalho, J.R. Henderson, and D.C. Hoyle, J. Chem. Phys. {\bf 100}, 591, (1994).

\bibitem{evans:86:0} R. Evans, U. Marini Bettolo Marconi, and P. Tarazona, J. Chem. Phys. {\bf 84}, 2376 (1986).

\bibitem{evans:87:0} R. Evans and U. Marini Bettolo Marconi, J. Chem. Phys. {\bf 86}, 7138 (1987).

\bibitem{marconi:88:0}  U. Marini Bettolo Marconi, Phys. Rev. A {\bf 38}, 6267 
(1988).

\bibitem{parry:92:0} A.O. Parry and R. Evans, Physica A {\bf 181}, 250 (1992).

\bibitem{krech:97:0} M. Krech, Phys. Rev. A {\bf 56}, 1642 (1997).

\bibitem{dietrich:98:0} A. Hanke, F. Schlesener, E. Eisenriegler, and S. Dietrich, Phys. Rev. Lett. {\bf 81}, 1885 (1998).

\bibitem{upton:99:0} Z. Borjan and P. Upton (unpublished); Z. Borjan, Ph.D. thesis, University of Bristol, 1999 (unpublished).

\bibitem{evans:94:0} R. Evans  and J. Stecki,  Phys.~Rev.~B {\bf 49}, 8842 
(1993).

\bibitem{fdg:78:0} M. E. Fisher and P.G. de Gennes, C. R. Acad Sci. Ser. B {\bf 287}, 207 (1978).

\bibitem{fss} for review see for example {\it Finite-Size Scaling and Numerical Simulations of Statistical Systems}, edited by V. Privman (World Scientific, Singapore, 1990).

\bibitem{au-yang:75:0} H. Au-Yang and M.E. Fisher, Phys. Rev. B {\bf 11},3469 (1975);  H. Au-Yang and M.E. Fisher, Phys. Rev. B. {\bf 21},3956 (1980).

\bibitem{abraham:71:0} D.B. Abraham, Stud. Appl. Math. {\bf 50}, 71 (1971).


\bibitem{abraham:73:0} D.B. Abraham and A. Martin-L{\"o}f, Commun. Math. Phys. {\bf 32}, 245 (1973).

\bibitem{maciolek:96:0} A. Macio\l ek, J. Phys. A {\bf 39}, 3837 (1996);
A. Macio\l ek and J. Stecki, Phys. Rev. B {\bf 54}, 1128 (1996).

                      
\bibitem{maciolek:99:0} A. Macio{\l}ek,  A. Drzewi{\' n}ski, and A. Ciach,  Phys. 
Rev. E {\bf 60}, 5009 (1999).


\bibitem{clsys} T. Nishino, J. Phys. Soc. Jpn. {\bf 64}, 3598 (1995).

\bibitem{carlon:98:0} E.~Carlon and A.~Drzewi\'nski, Phys. Rev. Lett. {\bf 79}, 1591 (1997).

\bibitem{drzew:00:1} A.~Macio\l ek, A.~Drzewi\'{n}ski, and R.~Evans, Phys. Rev. E {\bf 64}, 056137 (2001).

\bibitem{attard:91:0} P. Attard, D.R. B\'erard, C.P. Ursenbach, and G.N. Patey, Phys. Rev. A {\bf 44}, 8224 (1991).

\bibitem{krech:99:0} M. Krech, {\it The Casimir Effect in Critical System} (World Scientific, Singapore, 1994); J. Phys. Condens. Matter {\bf 11}, R391 (1999).

\bibitem{maciolek:01:0} A.~Macio\l ek, A.~Drzewi\'{n}ski, and R.~Evans,
Phys.~Rev.~E {\bf 64}, 056137 (2001). 

\bibitem{fisher:81:0} M.E. Fisher and H. Nakanishi,  J. Chem. Phys.
{\bf 75}, 5857 (1981); H. Nakanishi and M.E. Fisher,  J. Chem. Phys.
{\bf 78}, 3279 (1983).

\bibitem{frank:03:1} F. Schlesener, A. Hanke, and  S. Dietrich, J. Stat. Phys. {\bf 110}, 981 ( 2003). 

\bibitem{white} S.R. White, Phys. Rev. Lett. {\bf 69}, 2863 (1992);
S.R. White, Phys. Rev. B {\bf 48}, 10345 (1993).

\bibitem{dmrg} {\it Lectures Notes in Physics.} V.528
ed. by I. Peschel, X. Wang, M.Kaulke and K. Hallberg  (Springer, Berlin 1999).


\bibitem{bulk2} A.~Drzewi\'{n}ski, A.~Ciach, and A.~Macio\l ek, 
Eur.~Phys.~J.~B {\bf 5}, 825 (1998).

\bibitem{arnoldi} G.H. Golub and C.F. Van Loan, {\it Matrix Computations}
(Baltimore, 1996).


\bibitem{onsager:45:0} L. Onsager, Phys. Rev. {\bf 65}, 117 (1944).

\bibitem{diehl} H.W. Diehl, in {\it Phase Transitions and Critical Phenomena.} V. 10 p. 75 ed. by C. Domb and J.L. Lebowitz (Academic Press Inc., London, 1986).

\bibitem{dantchev} D. Dantchev, M. Krech, and S. Dietrich, Phys. Rev. E {\bf 67}, 066120 (2003).

\bibitem{blote:86:0} H. W. Bl{\"o}te, J. L. Cardy, and M. P. Nightingale, Phys. Rev. Lett. {\bf 56}, 742 (1986).


\bibitem{Weeks79} J.D. Weeks, D. Chandler, and H.C. Andersen, 
J. Chem. Phys. {\bf 54}, 5237 (1979).

\bibitem{Evans92} R. Evans,  in {\it Fundamentals of Inhomogeneous Fluids},
edited by D. Henderson, (Marcel, New York, 1992).

\bibitem{Rosenfeld89} Y. Rosenfeld, Phys. Rev. Lett. {\bf 63}, 980 (1989).

\bibitem{Rosenfeld93} Y. Rosenfeld, J. Chem. Phys. {\bf 98}, 8126 (1993).

\bibitem{Henderson92} J.R. Henderson,  in {\it Fundamentals of Inhomogeneous Fluids},
edited by D. Henderson, (Marcel, New York, 1992).

\bibitem{dijkstra:00:0} M. Dijkstra and R. Evans, J. Chem. Phys. {\bf 98}, 8126 (1993).


\bibitem{evans:94:2} R. Evans, Ber. Bunsenges. Phys. Chem. {\bf 112}, 1449 (2000).

\bibitem{grad} I. S. Gradshteyn and I. M. Ryzhik, {\it Table of Integrals, Series, and Products}, (Academic Press, New York, 1965).

\bibitem{footnote} Keeping all contribution  in  the Eq.~(\ref{eq:attard}) can lead to the result
that the total  solvation force is always attractive~\cite{attard:91:0}.

\bibitem{pertsin:03:0} A.J. Pertsin and M. Grunze,  J. Chem. Phys. {\bf 118}, 8004 (2003).

\end{thebibliography}
\end{document}